\documentclass[11pt]{article}
\usepackage{amsmath,amssymb,color,graphics,epsfig,cite,cancel,ulem}
\usepackage{amsfonts}
\newcommand{\hoch}[1]{$\, ^{#1}$}

\newcommand{\be}{\begin{equation}}
\newcommand{\ee}{\end{equation}}
\newcommand{\bea}{\setlength\arraycolsep{2pt} \begin{eqnarray}}
\newcommand{\eea}{\end{eqnarray}}
\newcommand{\nn}{\nonumber}

\def\ft#1#2{{\textstyle{\frac{\scriptstyle #1}{\scriptstyle #2} } }}

\def\0{{\sst{(0)}}}
\def\1{{\sst{(1)}}}
\def\2{{\sst{(2)}}}
\def\3{{\sst{(3)}}}
\def\4{{\sst{(4)}}}
\def\5{{\sst{(5)}}}
\def\6{{\sst{(6)}}}
\def\7{{\sst{(7)}}}
\def\8{{\sst{(8)}}}
\def\sst#1{{\scriptscriptstyle #1}}

\begin{document}

\begin{center}
{\Large {\bf Higher Derivative Contributions to Black Hole Thermodynamics at
NNLO}}

\vspace{20pt}

{\large Liang Ma\hoch{1}, Yi Pang\hoch{1} and H. L\"u\hoch{1,2}}

\vspace{10pt}

{\it \hoch{1}Center for Joint Quantum Studies and Department of Physics,\\
School of Science, Tianjin University, Tianjin 300350, China \\}

\bigskip

{\it \hoch{2}Joint School of National University of Singapore and Tianjin University,\\
International Campus of Tianjin University, Binhai New City, Fuzhou 350207, China}

\vspace{40pt}

\underline{ABSTRACT}

\end{center}

In an effective theory of gravity, thermodynamic quantities of black holes receive corrections from the infinite series of higher derivative terms. At the next to leading order, these can be obtained by using only the leading order solution. In this paper, we push forward this property to the next to next to leading order. We propose a formula which yields the Euclidean action of asymptotically flat black holes  at the next to next to leading order using only the solution up to and including the next to leading order. Other conserved quantities are derived from the Euclidean action via standard thermodynamic relation. We verify our formula in examples of $D$-dimensional pure gravity and Einstein-Maxwell theory extended by 4- and 6-derivative terms. Based on our formula, we also prove that for asymptotically flat black holes,  the physical quantities are invariant under field redefinitions.

\vfill{\footnotesize liangma@tju.edu.cn\ \ \ pangyi1@tju.edu.cn\ \ \ mrhonglu@gmail.com}
%\vfill {\footnotesize mrhonglu@gmail.com}

%{\footnotesize \hoch{*}Corresponding author}

\thispagestyle{empty}
\pagebreak

\tableofcontents
\addtocontents{toc}{\protect\setcounter{tocdepth}{2}}

\newpage

\section{Introduction}
It is widely believed that Einstein gravity coupled to traditional matter is not UV completed on its own. As suggested by string theory, the low energy effective action of a theory of quantum gravity should contain alongside the Einstein Hilbert term,  an infinite number of higher curvature terms suppressed by the scale $\Lambda_c$ where new degrees of freedom reside. Hence to look for signatures of new physics in the gravity sector beyond the standard model,
it is natural to consider extensions of Einstein gravity of the form
\be
I=\frac1{16\pi G}\int d^4x\sqrt{-g}\left(R+\Lambda_c^{-2}{\cal L}_{4\partial}+\Lambda_c^{-4}{\cal L}_{6\partial}+\cdots\right).\label{genlag0}
\ee
The higher-derivative gravity theory is defined as a perturbation theory and one solve it by
performing expansions of both fields and equations of motion in powers of a certain small parameter. In this approach, to compute a physical quantity at a required order in the small parameter,  one usually has to first find perturbed solutions at the same order. The standard perturbative approach has been applied to study higher curvature corrections to black holes \cite{Kats:2006xp, Cheung:2018cwt, Cano:2019ore, Cano:2019oma, Cremonini:2019wdk,Ma:2020xwi,  Agurto-Sepulveda:2022vvf }, black strings \cite{Ma:2021opb, Ma:2022nwq, Henriquez-Baez:2022bfi} and black branes \cite{Noumi:2022ybv}. This procedure however becomes increasingly complicated as the order increases.

It was first noticed in \cite{Reall:2019sah} that the next to leading order (NLO) contribution to thermodynamic quantities of an asymptotically flat black hole solution can be obtained without solving for the perturbed field equations at all. The leading order solution suffices. In practice, one only needs to evaluate the NLO higher-derivative action on the leading order solution. This approach has been successfully employed to obtain the thermodynamic quantities of black holes \cite{Loges:2019jzs, Cheung:2019cwi } and black strings  \cite{Ma:2021opb, Ma:2022nwq, Ma:2022gtm}. However, little is known beyond the NLO result.  Inspired by the work \cite{Reall:2019sah}, one expects that for asymptotically flat black holes,  knowing the perturbed solution up to $(n-1)$'th order is enough to derive the higher-derivative corrections to thermodynamic quantities up to and including the $n$'th order. Suppose this is true, it remains to be determined how one should proceed beyond the NLO.

In this work, we show that using the black hole solution with first order corrections
\be
\phi_{i}=\phi_{i(0)}+\epsilon \phi_{i(1)}\,,\label{allphifo}
\ee
where $\phi_i$ stands for all the fields in the theory, the Euclidean action up to and including order $\epsilon^2$ can be obtained by evaluating
\be
I_E[\phi_i]=I_0[\phi_{i(0)}]+\epsilon I_{hd1}[\phi_{i(0)}+\ft12\epsilon\phi_{i(1)}]+\epsilon^2I_{hd2}[\phi_{i(0)}]\,.
\label{onshell2}
\ee
Similar to \cite{Reall:2019sah}, in the pure gravity case, the quantity $T I_E$ of a static black hole corresponds to the free energy in the canonical ensemble with fixed temperature. In the Einstein-Maxwell case, $T I_E$ gives rise to the Gibbs free energy in the grand canonical ensemble where in addition to the temperature, the electrostatic potential is also fixed. This implies in the identification of $TI_E$ as the free energy of associated ensemble, one needs to perform a redefiniton of the integration constants in the solution such that  the temperature and electrostatic potential retain the same functional dependence on those constants. This is because in a typical method of deriving the perturbative solutions, one fixes either the horizon location or the conserved quantities such as mass and charges.

We confirm our proposal \eqref{onshell2} in two examples i) Einstein gravity with quadratic and cubic curvature terms; ii) Einstein-Maxwell theory extended by 4- and 6-derivative gauge-gravity couplings. Since the quadratic curvature terms in $D=4$ are topological up to field redefinitions, to verify our formula, the calculation is carried out in arbitrary spacetime dimensions. In $D=4$ pure gravity, the first nontrivial higher-derivative terms are  curvature cubed terms, thus the higher-derivative terms at the next order are quartic in curvature whose coefficients are not square of those of cubic curvature terms in their unit dimensions. (See \cite{Horowitz:2023xyl} for an example of this kind.) However, this is a rather special case. In the generic case, when matter couplings are turned on, the quadratic curvature terms can no longer be made topological. For instance, in the low energy effective action (in the Einstein frame) of heterotic string theory, the higher curvature terms are multiplied by scalar dependent functions.

Using our results, we prove the folklore that for asymptotically flat black holes, higher-derivative corrections to the thermodynamic quantities depend only on coefficients that are invariant under field redefinitions.

This paper is organized as follows. In section 2, we briefly review the derivation of \cite{Reall:2019sah} and show how it can be generalized to the next to next to leading order (NNLO). In section 3, built on the results from section 2, we prove that for asymptotically flat black holes the Euclidean action is unaffected by field redefinitions. In sections 4 and 5, we obtain the Euclidean actions of static black holes in 4- and 6-derivative extensions of pure gravity and Einstein-Maxwell theory using our formula (3) and show that they agree with those obtained by honestly solving the field equations perturbatively up to and including the second order in small parameters. We conclude in section 6.

\section{Generalizing Reall-Santos method to NNLO}
\subsection{A brief review of Reall-Santos method }
Consider a gravitational theory extended by higher derivative interactions characterized by a small parameter $\epsilon\ll 1$ in the unit of UV cutoff. In Euclidean signature, the action can be organized into the form
\be
I_E[\phi_i]=I_0[\phi_i]+\epsilon I_{hd}[\phi_i]\,,
\label{th1}
\ee
where $I_0$ is the 2-derivative action containing the Einstein-Hilbert term while
$I_{hd}$ denotes the leading order higher-derivative interactions. $I_0$ includes also the Gibbons-Hawking-York term \cite{York, Gibbons:1976ue} and a background subtraction term needed to yield a finite on-shell action for black hole solutions. All the field dependence of the action is conveniently denoted by $\phi_i$ where ``$i$'' labels different species.

It was noticed by Reall and Santos \cite{Reall:2019sah} that for asymptotically-flat black holes, the ${\cal O}(\epsilon)$ correction to their on-shell action can be neatly obtained by using {\it only} solutions to the leading order field equations, saving efforts from looking for the solution modified by the higher-derivative interactions. The proof goes as follows. We assume the solution with leading higher-derivative corrections to be the form
\be
\phi_{i}=\phi_{i(0)}+\epsilon \phi_{i(1)}\,,
\ee
where $\phi_{i(0)}$ and $\phi_{i(1)}$ denote the leading order solution and its leading order corrections respectively.
Substituting the solution to the action and expanding the result to the linear order in $\epsilon$, one obtains
\be
I_E[\phi_i]=I_0[\phi_{i(0)}]+\epsilon  I_{hd}[\phi_{i(0)}]+\epsilon\int d^Dx\,  \phi_{i(1)}(x)\,\frac{\delta I_0}{\delta \phi_{i}(x)}\Big|_{\phi_{i}=\phi_{i(0)}}\,,
\ee
where the last term vanishes since it is proportional to the leading order field equations. It was further shown that $\phi_{i(1)}$ does not contribute to the surface term in $I_0$, once the boundary conditions satisfied by $\phi_i$ remain fixed without ${\cal O}(\epsilon)$ corrections. Translated into the framework of black hole thermodynamics, fixing boundary conditions of fields amounts to fixing certain thermodynamic quantities determined by the thermodynamic ensemble. In the pure gravity example, it is the temperature
and angular velocity that remain fixed.  Thus one concludes that up to and including the first order in $\epsilon$, the on-shell action of asymptotically flat black hole solutions can be obtained by evaluating
\be
I_E[\phi_i]=I_0[\phi_{i(0)}]+\epsilon  I_{hd}[\phi_{i(0)}]\,,
\label{RS}
\ee
with certain thermodynamic quantities (determined by the thermodynamic ensemble) being fixed.
Surface terms generated during integration by part in recasting the bulk action into the form \eqref{RS} vanish because they fall off sufficiently fast at the infinity. The above procedure can be generalized to cases with multiplet independent small parameter characterizing the higher-derivative terms in the action.

\subsection{Generalization to include NNLO contribution}

In this section we generalize Reall-Santos method to models involving both leading and subleading higher-derivative corrections parameterized in the form
\be
I_E[\phi_i]=I_0[\phi_i]+\epsilon I_{hd1}[\phi_i]+\epsilon^2I_{hd2}[\phi_i]\,.
\label{th2}
\ee
Based on the result of Reall and Santos, it is conceivable that to obtain on-shell action of black hole solutions with corrections up to and including ${\cal O}(\epsilon^2)$, one needs only the solutions with corrections at ${\cal O}(\epsilon)$. There is no need to solve the pertubative field equations at  ${\cal O}(\epsilon^2)$. However, it is unclear what is the analogue of \eqref{RS} that one evaluates to obtain the on-shell action with ${\cal O}(\epsilon^2)$ corrections. We proceed as follows.

Suppose one has solved for the solutions up to and including ${\cal O}(\epsilon^2)$
\be
\phi_{i}=\phi_{i(0)}+\epsilon\phi_{i(1)}+\epsilon^2\phi_{i(2)}\,.
\label{nnlosol}
\ee
We substitute the equation above to \eqref{th2} and perform expansion up to order $\epsilon^3$
\bea
I_E[\phi_i]&=&I_0[\phi_{i(0)}]+\epsilon\Big(I_{hd1}[\phi_{i(0)}]+\int d^Dx \phi_{i(1)}(x)\frac{\delta I_0}{\delta \phi_{i}(x)}\Big|
_{\phi_{(i)}=\phi_{i(0)}}\Big)
\cr
&&+\epsilon^2 I_{hd2}[\phi_{i(0)}]+\epsilon^2\int d^Dx \Big(\phi_{i(1)}(x)\frac{\delta I_{hd1}}{\delta \phi_{i}(x)}\Big|
_{\phi_{i}=\phi_{i(0)}}+\phi_{i(2)}(x)\frac{\delta I_0}{\delta \phi_{i}(x)}\Big|
_{\phi_{i}=\phi_{i(0)}}\Big)
\cr
&&+\ft{1}{2}\epsilon^2\int d^Dx\int d^Dy\Big( \frac{\delta^2 I_0}{\delta \phi_{i}(x)\delta \phi_{j}(y)}\Big|
_{\phi_{i}=\phi_{i(0)}}\phi_{i(1)}(x)\phi_{j(1)}(y)\Big)+\cdots\,,
\label{I012}
\eea
in which terms proportional to the leading order field equations vanish. Notice that up to and including order $\epsilon$, the solution satisfies the field equations
{\it i.e.}
\be
\frac{\delta (I_0+\epsilon I_{hd1})}{\delta \phi_{i}(x)}\Big|_{\phi_{i}=\phi_{i(0)}+\epsilon\phi_{i(1)}}=0\,,
\ee
which leads to
\be
0=\int d^Dy\, \phi_{j(1)}(y)\,\frac{\delta^2 I_0}{\delta \phi_{j}(y)\delta \phi_{i}(x)}\Big|
_{\phi_{i}=\phi_{i(0)}}+\frac{\delta I_{hd1}}{\delta \phi_{i}(x)}\Big|
_{\phi_{i}=\phi_{i(0)}} \,.
\label{r1}
\ee
Using equation above in \eqref{I012}, we obtain the on-shell action of asymptotically-flat black holes up to and including the $\epsilon^2$ order:
\bea
I_E[\phi_i]&=&I_0[\phi_{i(0)}]+\epsilon I_{hd1}[\phi_{i(0)}]+\ft12\epsilon^2\int d^Dx\,
\phi_{i(1)}(x)\,\frac{\delta I_{hd1}}{\delta \phi_{i}(x)}\Big|
_{\phi_{i}=\phi_{i(0)}}+\epsilon^2 I_{hd2}[\phi_{i(0)}]\,,
\label{onshell1}
\eea
which can be further recast in the form \eqref{onshell2} given in the introduction. In order words, the second term in \eqref{onshell2} should be expanded in powers of $\epsilon$, and only terms of order $\epsilon$ and $\epsilon^2$ are kept. Following the same procedure as in \cite{Reall:2019sah}, we can show that when the boundary conditions imposed on various fields remain fixed, the boundary term contribution solely comes from $I_0$.

An immediate consequence is that the entropy correction with fixed mass and charges due to the higher-order terms \cite{Reall:2019sah}
\bea
\left(\frac{\partial S}{\partial\epsilon}\right)_{M,J,Q_e,\cdots}=-\left[\frac{\partial(\beta G)}{\partial\epsilon}\right]_{T,\Omega,\Psi_e,\cdots}\,,
\eea
up to and including the NNLO, is
\be
-\Delta S_{M,J,Q_e,\cdots}=\epsilon I_{hd1}[\phi_{i(0)}+\ft12\epsilon\phi_{i(1)}]+\epsilon^2I_{hd2}[\phi_{i(0)}]\,,
\label{entropy shift}
\ee
where $\phi_{i(1)}$ is the NLO perturbation that leaves $M, J, Q_e$, etc.~fixed. Expanding only to the $\epsilon$ order reduces to the result of \cite{Reall:2019sah}.

\section{Invariance of thermodynamics under field redefinitions}

In this section, we show that for asymptotically flat solutions, the on-shell action
evaluated using \eqref{onshell1} or \eqref{onshell2} is invariant under field redefinitions. Earlier attempts on proving this property was made in \cite{Jacobson:1993vj}. Recall that the on-shell action is related to the free energy via $G=T I_E$. Meanwhile temperature and other thermodynamic quantities which $G$ depends on are fixed as required by the boundary conditions. This can always be done in a black hole solution where the number of independent thermodynamic variables matches the number of parameters of the solution. Therefore, thermodynamic quantities derived from derivatives of $G$ enjoy the invariance too.

Consider arbitrary local field redefinitions up to and including ${\cal O}(\epsilon^2)$
\be
\phi_{i}\rightarrow\phi'_i=\phi_{i}+\epsilon{\cal F}_1[\phi_{i}]+\epsilon^2{\cal F}_2[\phi_{i}]\,,
\label{fred}
\ee
where ${\cal F}_1[\phi_{i}]$ and ${\cal F}_2[\phi_{i}]$ are local functions of $\phi_i$ and its derivatives.
Accordingly, the off-shell action is modified to
\be
I_E[\phi_i]\rightarrow I'_E[\phi_i]=I_{0}[\phi_i]+\epsilon I'_{hd1}[\phi_i]+\epsilon^2 I'_{hd2}[\phi_i]\,,
\ee
where
\bea
I'_{hd1}[\phi_i]&=&I_{hd1}[\phi_i]+\int d^Dx\, \frac{\delta I_0}{\delta \phi_i(x)}{\cal F}_1[\phi_{i}(x)]\,,
\cr
I'_{hd2}[\phi_i]&=&I_{hd2}[\phi_i]+\int d^Dx\, \frac{\delta I_0}{\delta \phi_i(x)}{\cal F}_2[\phi_{i}(x)]+\int d^Dx\, \frac{\delta I_{hd1}}{\delta \phi_{i}(x)}{\cal F}_1[\phi_{i}(x)]
\cr
&& +\ft{1}{2}\int d^Dx\int d^Dy\, \frac{\delta^2I_0}{\delta \phi_{i}(x)\delta \phi_{j}(y)}{\cal F}_1[\phi_{i}(x)]{\cal F}_1[\phi_{j}(y)]\,.
\eea
Given a solution to the original theory including all the higher-derivative corrections
\be
\phi_i=\phi_{i(0)}+\epsilon\phi_{i(1)}+\cdots\ ,
\ee
the field redefinition \eqref{fred} implies that the solution in the redefined theory becomes
\be
\phi'_{i}=\phi_{i(0)}+\epsilon(\phi_{i(1)}-{\cal F}_1[\phi_i])+\cdots\,.
\ee
Here we only write the leading order corrections explicitly because the higher order corrections are not needed in computing the on-shell action up to and including order $\epsilon^2$. The on-shell action evaluated using the redefined action is
\be
I'_{E}[\phi'_i]=I_0[\phi_0]+\epsilon I'_{hd1}[\phi_0+\ft12\epsilon(\phi_{i(1)}-{\cal F}_1[\phi_i])]+\epsilon^2 I'_{hd2}[\phi_0]\ .
\ee
At order $\epsilon$, it is easy to see that $I'_E[\phi'_i]=I_E[\phi_i]$ upon using the fact that $\delta I_0/\delta\phi_i(x)$ vanishes for $\phi_i=\phi_{i(0)}$. At order $\epsilon^2$, after a careful computation, the difference between $I'_E[\phi'_i]$ and $I_E[{\phi_i}]$ reduces to
\bea
&&I'_E(\phi'_i)- I_E(\phi_i)
\cr
&&=\ft12\epsilon^2\int d^Dx{\cal F}_1[\phi_{i(0)}(x)]\Big(\frac{\delta I_{hd1}}{\delta \phi_i(x)}\Big|_{\phi_i=\phi_{i(0)}}+\int d^Dy\phi_{j(1)}(y)\frac{\delta^2 I_0}{\delta\phi_i(x)\delta\phi_j(y)}\Big|_{\phi_i=\phi_{i(0)}}\Big),
\eea
which is again 0 using the identity \eqref{r1}. In the steps above, certain boundary terms
coming from integration by parts have been neglected because we can show that they fall off fast enough at the asymptotic infinity and do not contribute to the surface action. We thus conclude that for asymptotically flat black hole solutions, the on-shell action is invariant under field redefinitions.

In the next section, we apply our key formula \eqref{onshell1}, (or equivalently \eqref{onshell2},) to compute on-shell actions for spherically-symmetric and static black holes in pure gravity and Einstein-Maxwell theory extended by both 4- and 6-derivative terms in $D$-dimensions. We then confirm our results by directly evaluating the on-shell action for solutions that do contain the ${\cal O}(\epsilon^2)$ corrections.

\section{Example 1: static black hole in pure gravity}
\label{section static black hole}

In this section, we study thermodynamics of the Schwarzschild black hole in pure gravity extended by quadratic and cubic curvature invariants in arbitrary spacetime dimensions
\be
S=\frac{1}{16\pi G_D}\int d^Dx \sqrt{-g}\left(R+\mathcal{L}_{hd1}+\mathcal{L}_{hd2}\right)\,,
\label{pure gravity}
\ee
where the  quadratic and cubic curvature terms are given by
\bea
\label{quadratic term}
\mathcal{L}_{hd1}&=&\alpha_{1}R^{2}+\alpha_{2}R_{ab}R^{ab}+\alpha
_{3}R_{abcd}R^{abcd}\,,
\cr
\mathcal{L}_{hd2}&=&\beta_{1}R^{3}+\beta_{2}RR_{ab}R^{ab}+\beta
_{3}R_{b}^{a}R_{c}^{b}R_{a}^{c} +\beta_{4}R^{ab}R^{cd}R_{acbd}+ \beta_{5}RR_{abcd}R^{abcd}\cr
&&+\beta_{6} R^{ab}R_{acde}R_{b}^{\text{ }cde} +\beta_{7}R_{ab}^{\text{ \ }cd}R_{cd}^{\text{ \ }ef}R_{ef}^{\text{ \ }%
ab}+\beta_{8}R_{a\text{ }b}^{\text{ }c\text{ }d}R_{c\text{ }d}^{\text{
}e\text{ }f}R_{e\text{ }f}^{\text{ }a\text{ }b}\,,
\label{quadraction-cubic pure}
\eea
in which only $\alpha_3$, $\beta_7$, $\beta_8$ are invariant under the redefinition of the metric. The theory above is viewed as effective theory of gravity up to 6-derivative level. To be consistent with causality, it is natural to parameterize $\alpha_i$ and $\beta_i$ as $a_i\ell_D^2$ and $b_i\ell_D^4$ where $a_i, b_i$ are certain order 1 constant and $\ell_D$ is the $D$-dimensional Planck length. Thus the structure of these higher curvature corrections fits into the framework to which results from previous sections apply.

We adopt the most general ansatz for a spherically-symmetric and static metric
\bea
ds^{2}_{D} &=& -hdt^{2}+\frac{dr^2}{f} + r^2d\Omega^2_{D-2}\,.
\eea
From now on, we use $\omega_{D-2}=2\pi^{\frac{1}{2}(D-1)}/\Gamma[\frac{1}{2}(D-1)]$ to denote the volume of a $(D-2)$-dimensional unit sphere. We will first construct the NLO solution and then apply our formula \eqref{onshell1} (or equivalently \eqref{onshell2}) to obtain the Euclidean action and derive the free energy of the black hole involving NNLO contributions, namely the contributions from both the quadratic and cubic curvature terms. Other thermodynamic quantities simply follow from the standard thermodynamic relations.  Next we verify our results using the traditional methods by evaluating the on-shell action for the static black hole with NNLO higher derivative contributions.

\subsection{Our method}
In our method, to obtain Euclidean action black holes with both quadratic and cubic curvature corrections, we need only to find perturbative solutions modified by the quadratic curvature terms. Up to and including this order, we have
\be
h =h_0+\Delta h_1+\mathcal{O}(\alpha_i^2,\,\beta_i),\quad   f =f_0+\Delta f_1+\mathcal{O}(\alpha_i^2,\,\beta_i),\quad h_{0} =f_{0}=1-\frac{\mu}{r^{D-3}}
\label{perturb}
\ee
where the explicit expressions for corrections $\Delta h_1$ and $\Delta f_1$ are given in Appendix \eqref{cubic perturbed static solution}.  In the original unperturbed Schwarzschild  metric, there is only one integration constant $\mu$. The mass is proportional to $\mu$ and $r_0=\mu^{\frac{1}{D-3}}$ is the location of the black hole horizon.  In our specific construction of the perturbative solution, $\Delta h_1$ and $\Delta f_1$ have faster falloffs asymptotically, and hence the mass remains unchanged. However, the horizon is now shifted by
\be
r_h=r_0+\delta_1 r_h+\mathcal{O}(\alpha_i^2,\,\beta_i),\quad  \delta_1 r_h =-\frac{ (D-4) }{r_0}\alpha _3\ .
\label{r0shift}
\ee
The black hole temperature is also corrected
\be
T = \frac{(D-3) }{4 \pi r_0}(1-\frac{(D-4) (D-2) }{r_0^2}\alpha _3)+\mathcal{O}(\alpha_i^2,\,\beta_i)\ .
%%%
\label{temperature}
\ee
Note that only $\alpha_3$ appears in the temperature formula. This is because $\alpha_1$ and $\alpha_2$ are not invariant under the field redefinition and can be removed with an appropriate redefinition. As we have proven earlier, these coupling parameters will not enter the thermodynamic system in the perturbative approach with the mass or charges fixed.

Recall that to apply our formula one needs to fix certain thermodynamic quantities implied by the choice of ensemble. Here the thermodynamic ensemble is canonical ensemble with fixed temperature $T$. To achieve this, we need to perform a redefinition of the integration constant ${r}_0$ to $\tilde r_0$ so that up to and including the first-order corrections, the temperature retains the same functional form as the unperturbed solution with respect to $r_0$. We find that under
\be
r_0\rightarrow \tilde{r}_0+ \delta_1\tilde{r}_0+\mathcal{O}(\epsilon^2)\,,\qquad
\delta_1\tilde{r}_0=-\frac{ (D-4) (D-2) }{\tilde{r}_0}\alpha_3\,,
\label{r0 change}
\ee
the temperature takes the same form as the Schwarzschild black hole, namely
\be
T\rightarrow T_0=\frac{(D-3) }{4 \pi \tilde{r}_0}+\mathcal{O}(\alpha_i^2,\,\beta_i)\,.
\ee
After performing the shift of $\tilde{r}_0$ also in the solution, we can apply our formula to compute the Euclidean action of the temperature-fixed black hole. (In principle, we should drop the tilde on $r_0$ so that we get the leading-order perturbed solution of the Schwarzschild metric with fixed $T$; however, we shall retain the tilde symbol so as not to get confused.)

Specifically, we first substitute the $T$-fixed leading-order solution to $I_0$, $I_{hd1}$ and $I_{hd2}$, and we obtain
\bea
I_0[g_{ab(0)}]&=&\frac{\beta\omega_{D-2}}{16\pi}\tilde{r}_0^{D-3}\,,
\label{I0}\\
\epsilon I_{hd1}[g_{ab(0)}]&=&-\frac{\beta\omega_{D-2}}{16\pi} (D-3)(D-2)^2\tilde{r}_0^{D-5} \alpha_3\,,\label{I11}
\\
\epsilon^2 I_{hd2}[g_{ab(0)}]&=&-\frac{\beta\omega_{D-2}}{128\pi}(D-2)(D-3)\tilde{r}_0^{D-7}
\cr
&&\times \Big[4  \left(D^3-9 D^2+26 D-22\right)\beta _7+\left(3 D^2-15 D+16\right)\beta _8\Big],
\label{I2}
\eea
where $I_{hd1}[g_{ab(0)}]$ and $I_{hd2}[g_{ab(0)}]$ contain only contributions from the bulk action. However, $I_0$ includes the Gibbons-Hawking surface term and the background subtraction term
\be
I_0=-\frac{1}{16\pi}\int d^Dx \sqrt{g}R-\frac{1}{8\pi}\int d^{D-1}x\sqrt{h}\left(K-K_0\right)\,,
\ee
where the background metric appearing in the background subtraction term is
the flat metric
\be
ds^{2}_{D} = -h(r_c)dt^{2}+dr^2 + r^2d\Omega^2_{D-2}\,,
\ee
where $r_c$ is the location of the boundary hypersurface where the subtraction is performed. Eventually one sends $r_c$ to infinity.

Next, we substitute the $T$-fixed NLO perturbed solution, i.e.~\eqref{perturb} together with \eqref{r0 change}, into $I_{hd1}$ and expanding to ${\cal O}(\alpha_i^2)$. We find
\be
\epsilon I_{hd1}[g_{ab(0)}+g_{ab(1)} ]\big|_{{\cal O} (\alpha_i^2)}=\frac{\beta\omega_{D-2}}{16\pi} (D-4)^2(D-1)^2(D-3)(D-2)\tilde{r}_0^{D-7}\alpha_3^2\,,
\ee
which implies that the remaining term in our formula \eqref{onshell1} is given by
\be
\frac12\int d^Dx
g_{ab(1)}(x)\frac{\delta \epsilon I_{hd1}}{\delta g_{ab}(x)}\Big|
_{g_{ab}=g_{ab(0)}}=\frac{\beta\omega_{D-2}}{32\pi} (D-4)^2(D-3)(D-2)(D-1)^2\tilde{r}_0^{D-7}\alpha_3^2\ .
\label{I1}
\ee
Adding up \eqref{I0}, \eqref{I11}, \eqref{I2} and \eqref{I1}, we obtain the Euclidean action of the static black hole up to and including ${\cal O}(\alpha_i^2,\beta_i)$, given by
\bea
&&I_E(T_0,\alpha_i,\beta_i)=\frac{ \beta\omega _{D-2}}{16 \pi } \tilde{r}_0^{D-3}\Big[1-\frac{  (D-3) (D-2)^2 }{\tilde{r}_0^2}\alpha _3+\frac{(D-4)^2(D-3)(D-2)(D-1)^2}{2\tilde{r}_0^4}\alpha_3^2
\cr
&&\quad-\frac{(D-3) (D-2) }{8 \tilde{r}_0^4}\left(4  \left(D^3-9 D^2+26 D-22\right)\beta _7+ \left(3 D^2-15 D+16\right)\beta _8\right)
\Big],\label{It}
\eea
where $\tilde{r}_0$ is related to $T_0$ simply by $\tilde{r}_0=\frac{(D-3)}{ 4\pi T_0}$. Again, the on-shell action depends only on the coupling constants unaffected by the field redefinition. The Euclidean \eqref{It} leads to the Helmholtz free energy
\be
F(T_0,\alpha_i,\beta_i)=T_0 I_E(T_0,\alpha_i,\beta_i)\,,
\ee
from which one computes the energy and entropy of the black hole according to
\be
M=F(T_0,\alpha_i,\beta_i)+T_0 S\,,\qquad S=-\frac{\partial F(T_0,\alpha_i,\beta_i)}{\partial T_0}\,.
\ee
Explicit evaluation gives rise to
\bea
M&=&\frac{\left(D-2\right)\omega_{D-2}}{16\pi}\tilde{r}_0^{D-3}-\frac{\omega _{D-2}}{16 \pi } (D-4) (D-3) (D-2)^2 \tilde{r}_0^{D-5}\alpha _3\cr
&&+\frac{(D-6) (D-3) (D-2) \omega _{D-2}}{128 \pi } \Big[4  (D-4)^2 (D-1)^2\alpha _3^2-4  \left(D^3-9 D^2+26 D-22\right)\beta _7\cr
&&+ \left(-3 D^2+15 D-16\right)\beta _8\Big]\tilde{r}_0^{D-7}\,,
\cr
S &=&\frac{\omega_{D-2}}{4}\tilde{r}^{D-2}_{0}\Big[1-\frac{ (D-5) (D-2)^2}{\tilde{r}_0^2}\alpha _3+\frac{(D-7) (D-2)}{8 \tilde{r}_0^4}\big(4  (D-4)^2 (D-1)^2\alpha _3^2\cr
&&-4 (D^3-9 D^2+26 D-22)\beta _7 - (3 D^2-15 D+16)\beta _8
\big)\Big].
\eea
We therefore obtain the perturbed mass and entropy for fixed temperature, up to and including the NNLO perturbation, even though we have only made use of the NLO perturbed solution.

However, we are interested how the entropy is modified while the conserved quantity mass is fixed in the perturbation. In order to achieve this, we consider an appropriate redefinition of the integration constant $\tilde r_0$, expressed now in terms of $r_0$, namely
\bea
\tilde{r}_0&\rightarrow& r_0+ \delta_1r_0+\delta_2r_0\,,\qquad \delta_1r_0=\frac{ (D-4) (D-2) }{r_0}\alpha _3\,,
\cr
\delta_2r_0&=&\frac{ (D-6)}{8 r_0^3}\Big(4  \left(D^3-9 D^2+26 D-22\right)\beta _7+ \left(3 D^2-15 D+16\right)\beta _8\cr
&&-4  (D-4)^2 (2 D-3)\alpha _3^2
\Big),
\label{tilder0 and r0}
\eea
we obtain the all the thermodynamic variables with fixed mass
\bea
M'&=&\frac{\omega_{D-2}(D-2)}{16\pi}r_0^{D-3}\,,\cr
%%%
T' &=& \frac{(D-3) }{4 \pi r_0}\Big[1-\frac{(D-4) (D-2) }{r_0^2}\alpha _3+\frac{(D-4)^2 (4 D^2-23 D+26) }{2 r_0^4}\alpha _3^2
\cr
&&-\frac{(D-6) }{8 r_0^4}\left(4  (D^3-9 D^2+26 D-22)\beta _7+ (3 D^2-15 D+16)\beta _8 \right)\Big]\,,\cr
S' &=&\frac{\omega_{D-2}}{4}r^{D-2}_{0}\Big[1+\frac{ (D-2)^2 }{r_0^2}\alpha _3-\frac{(D-4)^2 (D-2) (2 D-3) }{2 r_0^4}\alpha _3^2\cr
&&+\frac{(D-2) }{8 r_0^4}\left(4  \left(D^3-9 D^2+26 D-22\right)\beta _7+\left(3 D^2-15 D+16\right)\beta _8 \right)
\Big]\,,
\label{thermodynamic fix mass}
\eea
Here we used ``prime" to label the mass fixed case involving the second-order perturbations.
Note that as one should have expected, $T'$ at the leading-order perturbation, namely the $\alpha_3$ term,  precisely recover the temperature in \eqref{temperature}. The Euclidean action associated with the fixed $M$ is
\bea
I'_E(T')&=&\beta(M'-T'S')=\frac{  \beta\omega _{D-2}}{16 \pi }r_0^{D-3}\cr
&&\times\Big[1-\frac{2  (D-3) (D-2) }{r_0^2}\alpha _3+\frac{2  (D-4) (D-3) \left(3 D^2-15 D+16\right)}{r_0^4}\alpha _3^2\cr
&&-\frac{(D-3) }{2 r_0^4}\left(4 \left(D^3-9 D^2+26 D-22\right) \beta _7+\left(3 D^2-15 D+16\right) \beta _8\right)\Big].
\label{It2}
\eea
As a consistent check, we can also obtain $I'_E(T')$ from \eqref{It} by substituting \eqref{tilder0 and r0} directly into \eqref{It}. Finally, we can check the entropy shift formula \eqref{entropy shift}. The corrected entropy \eqref{thermodynamic fix mass} and corrected Euclidean action \eqref{It}
\bea
\Delta S'&=&\frac{(D-2)\omega_{D-2}}{4}r^{D-2}_{0}\Big[\frac{ (D-2) }{r_0^2}\alpha _3-\frac{(D-4)^2  (2 D-3) }{2 r_0^4}\alpha _3^2\cr
&&+\frac{1 }{8 r_0^4}\left(4  \left(D^3-9 D^2+26 D-22\right)\beta _7+\left(3 D^2-15 D+16\right)\beta _8 \right)
\Big],\cr
\Delta I_E&=&-\frac{ (D-2)\omega _{D-2}}{4} \tilde{r}_0^{D-2}\Big[\frac{  (D-2) }{\tilde{r}_0^2}\alpha _3-\frac{(D-4)^2(D-1)^2}{2\tilde{r}_0^4}\alpha_3^2
\cr
&&+\frac{ 1 }{8 \tilde{r}_0^4}\left(4  \left(D^3-9 D^2+26 D-22\right)\beta _7+ \left(3 D^2-15 D+16\right)\beta _8\right)
\Big],
\eea
satisfy the formula \eqref{entropy shift} under the parameter relation \eqref{tilder0 and r0}. Specifically, in applying \eqref{entropy shift} appropriately, we need to involve the parameter redefinition such that
\bea
\Delta S'\Big|_{M_0} &=& -\Big(\Delta I_E\Big|_{T_0}\Big)\Big|_{\tilde r_0\rightarrow r_0+ \delta_1r_0}\nn\\
&=& -\Big(\epsilon I_{hd1}[\phi_{i(0)}+\ft12\epsilon\phi_{i(1)}]+\epsilon^2I_{hd2}[\phi_{i(0)}]
\Big)\Big|_{M_0}\,.\label{subtlety}
\eea
In other words, in order to obtain all the thermodynamic variables to the NNLO, we need to navigate back and forth between the original $M$-fixed and $T$-fixed perturbations. However, if we are only interested in $\Delta S'$ with fixed $M$ to NNLO, we can simply plug the $M$-fixed NLO perturbation \eqref{perturb} into the last step of \eqref{subtlety}, thereby simplifying the calculation significantly.

\subsection{Poor man's approach by solving second-order perturbative solutions}

In the previous section, we constructed the mass-fixed solution \eqref{perturb} up to and including NLO perturbation; however, after a series manipulation, we obtain the perturbed temperature and entropy \eqref{thermodynamic fix mass} up to and including the NNLO. In this section, we verify this result by solving explicitly the perturbed black hole solution to the NNLO and using it in the evaluation of the on-shell Euclidean action directly.

The mass-fixed static black hole with corrections from both the quadratic and cubic curvature terms is given in \eqref{cubic perturbed static solution} in Appendix, where $r_0$ denotes the horizon radius of the uncorrected Schwarzschild black hole.  With the higher-order contributions, we find the horizon radius is modified to
\bea
r_h&=&r_0+\delta_1 r_{h}+\delta_2 r_{h},\quad
\delta_1 r_{h} =-\frac{ (D-4) }{r_0}\alpha _3\,,
\cr
\delta_2 r_{h} &=& -\frac{1}{8r_0^3}\Big[8(D-3) (D-2) (D-1) \beta _5+4(D-3)^2 (D-1) \beta _6
\cr
&&+4(5 D^3-39 D^2+100 D-86)\beta _7+(3 D^2-21 D+38)\beta _8
\cr
&&-4 \alpha _3 (D-4)\left( 4 (D-1)(D-3) (\alpha _1+\alpha _2) +(13 D^2-58 D+60)\alpha _3\right)
\Big].
\eea
Having obtained this cumbersome solution, it is straightforward to  verify that the thermodynamic quantities associated with the mass-fixed perturbed solution are indeed the ones given in \eqref{thermodynamic fix mass}. The verification is straightforward albeit tedious and we shall not give the details here.

\section{Example 2: static black hole in Einstein-Maxwell gravity}

In this section, we consider 4-and 6-derivative corrections to charged static black holes in $D$-dimensional Einstein-Maxwell theory. The action of the theory begins with
\be
S_{EM}=\int d^Dx\sqrt{-g}\left[\frac1{16\pi G_D}R-\frac{F^2}{4e^2_D}+\cdots \right],
\ee
where the ellipsis represents the higher-derivative operators generated by integrating out massive degrees of freedom from energy scale near the Planck mass $M_p$. Before carrying out the perturbative calculation, one must understand the order of parameters in front of the higher derivative terms. This can be done by performing the rescaling of the Maxwell field
\be
F_{\mu\nu}\rightarrow \frac{e_D}{\sqrt{16\pi G_D}} F_{\mu\nu}\,.
\ee
We then define the dimensionless gauge coupling constant $g$ which is relate to $e_D$
via $e^2_D=g^2 M_p^{4-D}$.
After this is done, we present the model in the form below
\bea
{\cal S}_{EM}&=&\frac1{16\pi G_D}\int d^Dx\sqrt{-g}\Big[R-\frac{F^2}{4}+{\cal L}_{hd1}+{\cal L}_{hd2}\Big],\nn\\
\mathcal{L}_{hd1}&=&\alpha_3R_{abcd}R^{abcd}+\alpha_4R_{abcd}F^{ab}F^{cd}+\alpha_5(F_{ab}F^{ab})^2+\alpha_6F_a^{\ b}F_b^{\ c}F_c^{\ d}F_d^{\ a}\,,
\cr
\mathcal{L}_{hd2}&=&\beta_{7}R_{ab}^{\text{ \ }cd}R_{cd}^{\text{ \ }ef}R_{ef}^{\text{ \ }%
ab}+\beta_{8}R_{a\text{ }b}^{\text{ }c\text{ }d}R_{c\text{ }d}^{\text{
}e\text{ }f}R_{e\text{ }f}^{\text{ }a\text{ }b}+\gamma_{1}F_a^{\ c}F^{ab}F_b^{\ d}F^{ef}R_{cdef}\cr
&&+\gamma_{2}F_a^{\ c}F^{ab}F_d^{\ f}F^{de}R_{becf}+\gamma_{3}R_{abcd}R^{abce}F^{df}F_{ef}+\gamma_{4}R_{abcd}R^{abef}F^c_{\ e}F^d_{\ f}\,,
\label{Einstein-Maxwell}
\eea
where the orders of various parameters are
\bea
&&\alpha_3\sim{\cal O}(\frac1{M_p^2})\,,\qquad \alpha_4\sim{\cal O}(\frac{g^2}{M_p^2})\,,\qquad \alpha_5,\,\alpha_6\sim {\cal O}(\frac{g^4}{M_p^2})~~{\rm or}~~{\cal O}(\frac{g^2}{M_p^2})\,,
\cr
&&\beta_7,\, \beta_8 \sim {\cal O}(\frac1{M_p^4})\,, \qquad \gamma_1,\gamma_2\sim {\cal O}(\frac{g^4}{M_p^4})\,,\qquad \gamma_3,\gamma_4\sim {\cal O}(\frac{g^2}{M_p^4})\,.
\eea
Notice that the coefficients of $F^4$ terms can be of order ${\cal O}(\frac{g^4}{M_p^2})$ or ${\cal O}(\frac{g^2}{M_p^2})$ because when writing down the effective action above, we have chosen a frame in which certain $RF^2$ terms have been removed by field redefinition. For instance,  a term $\frac{g^2}{M_p^2} R_{a}^{b}F^{ac}F_{bc}$ in the original theory can be removed by redefining the metric. Consequently certain $F^4$ terms with coefficient of ${\cal O}(\frac{g^2}{M_p^2})$ are generated. For an example of this kind, one can see \cite{Alberte:2020bdz}. For $g\ll 1$, one observes that there exits a hierarchy among the small parameters since
\be
\frac{g^4}{M_p^2}\ll \frac{g^2}{M_p^2}\ll \frac{1}{M_p^2}\,.
\ee
Thus if $\alpha_5\,,\alpha_6$ are at ${\cal O}(\frac{g^2}{M_p^2})$, we see that the effective action is controlled by two independent parameters say  $\alpha_3$ and $\alpha_4$ and
\be
\beta_7,\beta_9\sim \alpha_3^2\,\qquad\gamma_1,\gamma_2\sim\alpha_4^2\,,\qquad \gamma_3,\gamma_4\sim \alpha_3\alpha_4\,.
\ee
If the original theory does not have $R_{a}^{b}F^{ac}F_{bc}$ and $R F^2$, then $\alpha_5$ and $\alpha_6$ are at ${\cal O}(\frac{g^4}{M_p^2})$ which will be the third independent parameter. In the computation, we do not distinguish the two cases for $\alpha_5$ and $\alpha_6$. The results are computed up to terms proportional to $\alpha_i^2,\,\beta_i$ and $\gamma_i$.

\subsection{Our method}
To apply our method, we only need to solve the black hole solution corrected by the 4-derivative terms. We find that the static charged black hole is modified to
\bea
ds^{2}_{D}&=&-hdt^{2}+\frac{dr^2}{f} + r^2d\Omega^2_{D-2},\qquad A_{(1)}=\psi(r)dt\,,
\cr
h&=&h_0+\Delta h_1+\cdots,\qquad   f =f_0+\Delta f_1+\cdots,\qquad \psi =\psi_0+\Delta \psi_1+\cdots\,,\nn\\
%%%
h_0&=&f_0=1-\frac{\mu }{r^{D-3}}+\frac{q^2}{r^{2 (D-3)}},\qquad \psi_0= \sqrt{\frac{2(D-2)}{D-3}}\frac{ q}{r^{D-3}}\,,
\label{charge solution}
\eea
where the correction terms are of the order ${\cal O}(\alpha_i)$. Their explicit form are given in \eqref{Maxwell quadratic} in the Appendix. The unperturbed solution has two independent parameters $(\mu,q)$, parameterizing the mass and charge respectively. In this perturbation \eqref{Maxwell quadratic}, the mass and the charge are fixed, but the horizon is shifted. The horizon of the unperturbed solution is denoted by $r_0$, satisfying
\be
\mu= r_0^{D-3}+\frac{q^2}{r_0^{D-3}}\,.
\ee
With the 4-derivative corrections, the radius of horizon is shifted to
\bea
r_h&=&r_0+\delta_1 r_h+{\cal O}(\alpha_i^2)\,,
\cr
\delta_1 r_h &=&-\frac{1}{(D-2) (3 D-7) ( r_0^{2 D}-q^2 r_0^6)r_0^{2 D+1}}\Big[
\Big((D-4) (D-2) (3 D-7)  r_0^{4 D}\cr
&&-2 (D-3) (3 D-8) (3 D-7)  q^2 r_0^{2 D+6}+(D-3) (D-2) (11 D-32) q^4 r_0^{12}
\Big)\alpha_3\cr
&&+2 (D-3) (D-2) q^2 r_0^6\Big((D-1) q^2 r_0^6-(3 D-7)  r_0^{2 D}
\Big)\alpha_4\cr
&&-8  (D-3) (D-2)^2 q^4 r_0^{12}(2 \alpha _5+\alpha _6)
\Big].\label{horizon shift 0}
\eea
Furthermore, both temperature and electrostatic potential are modified, giving
\bea
&&T=T_0+\Delta T_1+\mathcal{O}(\alpha_i^2),\qquad \Psi_e=\Psi_{e,0}+\Delta \Psi_{e,1}+\mathcal{O}(\alpha_i^2)\,,
\cr
&&T_0=\frac{(D-3) }{4 \pi  r_0^{2 D-5}}( r_0^{2D-6}-q^2),\qquad \Psi_{e,0}=\sqrt{\frac{2(D-2)}{D-3}}\frac{ q}{r_0^{D-3}}\,,
\cr
&&\Delta T_1=-\frac{(D-3) r_0^{-4 D-3}}{4 \pi  (D-2) (3 D-7) ( r_0^{2 D}-q^2 r_0^6)}\Big[
\Big((D-3) (D-2)(7 D^2-36 D+48) q^6 r_0^{18}\cr
&&~+(D-4) (D-2)^2 (3 D-7)  r_0^{6 D}-(D-3) (15 D^3-114 D^2+300 D-272)  q^4 r_0^{2 (D+6)}\cr
&&~-(D-4) (D-2)^2 (3 D-7)  q^2 r_0^{4 D+6}
\Big)\alpha_3+2 (D-3) (D-2) q^2 r_0^6\Big((D-2) (5 D-13) q^4 r_0^{12}\cr
&&~+(D-2) (3 D-7)  r_0^{4 D}-4 (3 D^2-15 D+19)  q^2 r_0^{2 D+6}
\Big)\alpha_4\cr
&&~-8  (D-3) (D-2)^2 q^4 r_0^{12} ((2-D) q^2 r_0^6+(3 D-8) r_0^{2 D})(2 \alpha _5+\alpha _6)
\Big],
\cr
%%%
&&\Delta\Psi_{e,1}=\frac{\sqrt{2} \sqrt{D-3} q r_0^{1-3 D}}{\sqrt{D-2} (3 D-7)( r_0^{2 D}-q^2 r_0^6)}\Big[\Big((D-3) (7 D^2-36 D+48) q^4 r_0^{12}\cr
&&~+(D-4) (D-2) (3 D-7)  r_0^{4 D}-2 (D-3) (7 D^2-36 D+48)  q^2 r_0^{2 D+6}
\Big)\alpha_3\cr
&&~+2 (D-3) (D-2)\Big((3 D-7) r_0^{4 D}-2 (5 D-13)  q^2 r_0^{2 D+6}+(5 D-13) q^4 r_0^{12}
\Big)\alpha_4\cr
&&~+8  (D-3) (D-2)^2 q^2 r_0^6 (q^2 r_0^6-2 r_0^{2 D})(2 \alpha _5+\alpha _6)
\Big].
\label{fix mass and charge}
\eea
Here all the thermodynamic quantities are expressed as functions of integration constants $(r_0, q)$. Next, as required by our method, we redefine the integration constants $r_0$ and $q$ such that the functional dependence of temperature and the electrostatic potential on $r_0$ and $q$ are not modified up to and including the first order in $\alpha_i$. This is because the ensemble of the formalism specified by the action \eqref{Einstein-Maxwell} requires that temperature and electrostatic potential be fixed. We find
\bea
&&r_0\rightarrow \tilde{r}_0+ \delta_1\tilde{r}_0,\quad q\rightarrow \tilde{q}+ \delta_1\tilde{q},\cr
&&\delta_1\tilde{r}_0=-\frac{(D-4) \tilde{r}_0^{-2 D-1}}{(D-2) (3 D-7) \left( \tilde{r}_0^{2 D}-\tilde{q}^2 \tilde{r}_0^6\right)}\Big[\big(
(D-3) \left(7 D^2-36 D+48\right) \tilde{q}^4 \tilde{r}_0^{12}\cr
&&\ \ +(D-2)^2 (3 D-7)  \tilde{r}_0^{4 D}-2 (D-3) (D-2) (3 D-7)  \tilde{q}^2 \tilde{r}_0^{2 D+6}
\big)\alpha_3\cr
&&\ \ +2 (D-3) (D-2) \tilde{q}^2 \tilde{r}_0^6  \left((5 D-13) \tilde{q}^2 \tilde{r}_0^6-(3 D-7)  \tilde{r}_0^{2 D}\right)\alpha_4\cr
&&\ \ +8 (D-3) (D-2)^2 \tilde{q}^4 \tilde{r}_0^{12}\left(2 \alpha _5+\alpha _6\right)
\Big],\cr
&&\ \ \delta_1\tilde{q}=-\frac{(D-3) \tilde{q} \tilde{r}_0^{-2 D-2}}{(D-2) (3 D-7) \left( \tilde{r}_0^{2 D}-\tilde{q}^2 \tilde{r}_0^6\right)}\Big[\big(-2 (D-3) \left(3 D^3-18 D^2+30 D-8\right)  \tilde{q}^2 \tilde{r}_0^{2 D+6}\cr
&&\ \ +(D-3)^2 \left(7 D^2-36 D+48\right) \tilde{q}^4 \tilde{r}_0^{12}+(D-4) (D-2) (D-1) (3 D-7)  \tilde{r}_0^{4 D}
\big)\alpha_3\cr
&&\ \ +2 (D-3) (D-2)\big((3 D-7)  \tilde{r}_0^{4 D}+(D-3) (5 D-13) \tilde{q}^4 \tilde{r}_0^{12}-\left(3 D^2-9 D+2\right)  \tilde{q}^2 \tilde{r}_0^{2 D+6}
\big)\alpha_4\cr
&&\ \ -8  (D-3) (D-2)^2 \tilde{q}^2 \tilde{r}_0^6 \left(-D \tilde{q}^2 \tilde{r}_0^6+2 \tilde{r}_0^{2 D}+3 \tilde{q}^2 \tilde{r}_0^6\right)\left(2 \alpha _5+\alpha _6\right)
\Big].
\label{rq transformation0}
\eea
We can check that indeed with the above redefiniton in terms of $\tilde{r}_0$ and $\tilde{q}$, both the temperature $T$ and potential $\Psi_e$ retain their functional forms as the leading order expressions, namely
\be
T=T_0+\mathcal{O}(\alpha_i^2),\quad \Psi_e=\Psi_{e,0}+\mathcal{O}(\alpha_i^2)\ .
\ee
Plugging the leading-order solution into the action, we have
\bea
I_0[\phi_{i(0)}]&=& \frac{\beta\omega_{D-2}}{16\pi}(\tilde{r}_0^{D-3}-\frac{\tilde{q}^2}{\tilde{r}_0^{D-3}})\,,
\cr
\epsilon I_{hd1}[\phi_{i(0)}]&=&-\frac{\beta\omega_{D-2}(D-3)}{16\pi(3 D-7) \tilde{r}_0^{3 D+5}}\Big[\Big((D-3) (7 D^2-36 D+48) \tilde{q}^4 \tilde{r}_0^{12}+(D-2)^2 (3 D-7)  \tilde{r}_0^{4 D}\cr
&&-2 (D-3) (D-2) (3 D-7)  \tilde{q}^2 \tilde{r}_0^{2 D+6}
\Big)\alpha_3+2 (D-3) (D-2) \tilde{q}^2 \tilde{r}_0^6 \big((5 D-13) \tilde{q}^2 \tilde{r}_0^6\cr
&&-(3 D-7)  \tilde{r}_0^{2 D}\big)\alpha_4+8  (D-3) (D-2)^2 \tilde{q}^4 \tilde{r}_0^{12}(2 \alpha _5+\alpha _6)
\Big]\,,
\cr
%%%
\epsilon^2 I_{hd2}[\phi_{i(0)}]&=&\frac{\beta\omega_{D-2}(D-3) \tilde{r}_0^{-5 D-7}}{128\pi (3 D-5) (5 D-11)}\big[I_{2,\beta_7}\beta_{7}+I_{2,\beta_8}\beta_{8}+I_{2,2 \gamma_{1}+\gamma_{2}}(2 \gamma_{1}+\gamma_{2})\cr
&&+I_{2,\gamma_{3}}\gamma_{3}+I_{2,\gamma_{4}}\gamma_{4}
\big]\,,
\label{action1}
\eea
where to shorten the expression we have defined
\bea
I_{2,\beta_7}&=&4\Big[
3 (D-3) (D-2) (5 D-11) \left(5 D^3-37 D^2+86 D-62\right)  \tilde{q}^2 \tilde{r}_0^{4 D+6} \cr
&&-(D-2) (3 D-5)(5 D-11) \left(D^3-9 D^2+26 D-22\right)  \tilde{r}_0^{6 D}\cr
&&-3 (D-3)^2 (D-2) (5 D-11) \left(11 D^2-54 D+62\right)  \tilde{q}^4 \tilde{r}_0^{2 (D+6)}\cr
&&+(D-3)^2 (129 D^4-1239 D^3+4410 D^2-6886 D+3964) \tilde{q}^6 \tilde{r}_0^{18}
\Big]\,,
\nn\\
I_{2,\beta_8}&=&(D-2)\Big[3 (D-3) (5 D-11) \left(7 D^2-31 D+32\right)  \tilde{q}^2 \tilde{r}_0^{4 D+6}\cr
&&-(3 D-5) (5 D-11) \left(3 D^2-15 D+16\right)  \tilde{r}_0^{6 D}+3 (D-3)^2 (11 D^2\cr
&&-45 D+44) \tilde{q}^6 \tilde{r}_0^{18}-3 (D-3)^2 (5 D-11) (7 D-12)  \tilde{q}^4 \tilde{r}_0^{2 (D+6)}
\Big]\,,
\nn\\
I_{2,2 \gamma_{1}+\gamma_{2}}&=&8 (D-3)^2 (D-2)^2 (3 D-5) \tilde{q}^4 \tilde{r}_0^{12} \Big[(11 D-29) \tilde{q}^2 \tilde{r}_0^6-(5 D-11)  \tilde{r}_0^{2 D}\Big]\,,
\nn\\
I_{2,\gamma_{3}}&=&16 (D-3)^2 (D-2) \tilde{q}^2 \tilde{r}_0^6\Big[
(23 D^3-158 D^2+357 D-262) \tilde{q}^4 \tilde{r}_0^{12}\cr
&&+(D-2) (D-1) (5 D-11)  \tilde{r}_0^{4 D}-4 (D-2)^2 (5 D-11)  \tilde{q}^2 \tilde{r}_0^{2 D+6}
\Big]\,,
\nn\\
I_{2,\gamma_{4}}&=&16 (D-3)^2 (D-2) \tilde{q}^2 \tilde{r}_0^6\Big[\left(23 D^3-160 D^2+364 D-269\right) \tilde{q}^4 \tilde{r}_0^{12}\cr
&&+(D-2)^2 (5 D-11)  \tilde{r}_0^{4 D}-(5 D-11) \left(4 D^2-17 D+17\right)  \tilde{q}^2 \tilde{r}_0^{2 D+6}
\Big].
\eea
Similar to the uncharged case, upon substituting the $(T,\Psi)$-fixed solution with 4-derivative corrections into the 4-derivative action and evaluating the action,  we can read off the ${\cal O}(\alpha_i^2)$ contributions, which lead to
\bea
&&\frac{1}{2}\int d^Dx
\epsilon_i\phi_{i(1)}(x)\frac{\delta \epsilon I_{hd1}}{\delta \phi_i(x)}\Big|
_{\phi_i=\phi_{i(0)}}\cr
=&&\frac{\beta\omega_{D-2}(D-3) \tilde{r}_0^{-5 D-7}(
I_{1,\alpha_3^2}\alpha_3^2+I_{1,\alpha_3\alpha_4}\alpha_3\alpha_4+I_{1,\alpha^2_4}\alpha_4^2)}{32\pi (D-2) (3 D-7)^2 (3 D-5) (5 D-11) \left( \tilde{r}_0^{2 D}-\tilde{q}^2 \tilde{r}_0^6\right)}
\nn\\
&&+\frac{\beta\omega_{D-2}  (D-3)^2 (D-2) \tilde{q}^4 \tilde{r}_0^{5-5 D}(2 \alpha _5+\alpha _6)}{2\pi(3 D-7)^2 (5 D-11) (\tilde{r}_0^{2 D}-\tilde{q}^2 \tilde{r}_0^6)}\Big[I_{1,(2 \alpha _5+\alpha _6) \alpha _3} \alpha _3
\nn\\
&&
+2 (D-3) (D-2)I_{1,(2 \alpha _5+\alpha _6) \alpha _4}\alpha _4+4 (D-3) (D-2)^2 \tilde{q}^2 \tilde{r}_0^6I_{1,\left(2 \alpha _5+\alpha _6\right)^2}\left(2 \alpha _5+\alpha _6\right)
\Big],
\label{action2}
\eea
where to shorten the expression, we have defined
\bea
&&I_{1,\alpha_3^2}=\Big[(D-3)^2 (735 D^8-14716 D^7+128658 D^6-641968 D^5+2000407 D^4\cr
&&~-3986580 D^3+4961176 D^2-3523424 D+1092608) \tilde{q}^8 \tilde{r}_0^{24}\cr
&&~+(D-4)^2 (D-2)^2 (D-1)^2 (3 D-7)^2 (3 D-5) (5 D-11)  \tilde{r}_0^{8 D}\cr
&&~+2 (D-3) (D-2) (3 D-7) (5 D-11) (39 D^6-533 D^5+2931 D^4-8255 D^3+12482 D^2\cr
&&~-9592 D+3008)  \tilde{q}^4 \tilde{r}_0^{4 (D+3)}-4 (D-3)^2 (315 D^8-5649 D^7+43457 D^6-186397 D^5\cr
&&~+483704 D^4-767058 D^3+707276 D^2-326928 D+47872)  \tilde{q}^6 \tilde{r}_0^{2 (D+9)}\cr
&&~-4 (D-4) (D-3) (D-2)^2 (D-1) (3 D-7)^2 (5 D-11) (3 D^2-10 D+4)  \tilde{q}^2 \tilde{r}_0^{6 D+6}
\Big]\,,
\nn\\
&&I_{1,\alpha_3\alpha_4}=4 (D-3) (D-2) \tilde{q}^2 \tilde{r}_0^6\Big[(D-3) (525 D^7-9944 D^6+79474 D^5-348200 D^4
\cr
&&~+904589 D^3-1394776 D^2+1182404 D-425144) \tilde{q}^6 \tilde{r}_0^{18}-(D-3) (765 D^7-13242 D^6
\cr
&&~+95788 D^5-375134 D^4 +857419 D^3-1140104 D^2+812732 D-237968)  \tilde{q}^4 \tilde{r}_0^{2 (D+6)}
\cr
&&~+(3 D-7) (5 D-11) (33 D^6-487 D^5+2873 D^4-8601 D^3+13610 D^2-10572 D
\cr
&&~+3032)  \tilde{q}^2 \tilde{r}_0^{4 D+6} -(D-4) (D-2) (D-1) (3 D-7)^2 (5 D-11) \left(3 D^2-15 D+16\right)  \tilde{r}_0^{6 D}\Big]\,,
\nn\\
&&I_{1,\alpha^2_4}=4 (D-3)^2 (D-2)^2 \tilde{q}^2 \tilde{r}_0^6\Big[(375 D^6-7399 D^5+55738 D^4-211970 D^3+435991 D^2
\cr
&&\quad -463767 D+200248) \tilde{q}^6 \tilde{r}_0^{18}+2 (D-1)^2 (3 D-7)^2 (5 D-11)  \tilde{r}_0^{6 D}
\cr
&&\quad -2 (3 D-5) \left(75 D^5-1349 D^4+8534 D^3-25118 D^2+35255 D-19125\right)  \tilde{q}^4 \tilde{r}_0^{2 (D+6)}
\cr
&&\quad +(3 D-7) (3 D-5) (5 D-11) \left(3 D^3-42 D^2+131 D-108\right)  \tilde{q}^2 \tilde{r}_0^{4 D+6}
\Big]\,,
\nn\\
&&I_{1,\left(2 \alpha _5+\alpha _6\right) \alpha _3}=
2 (D-3) (-15 D^5+210 D^4-1069 D^3+2448 D^2-2398 D+664) \tilde{q}^2 \tilde{r}_0^{2 D+6}
\cr
&&\quad +(D-3)(35 D^5-561 D^4+3429 D^3-10147 D^2+14684 D-8368) \tilde{q}^4 \tilde{r}_0^{12}
\cr
&&~+(D-5) (D-4) (D-2) (D-1) (3 D-7) (5 D-11) \tilde{r}_0^{4 D}\,,
\nn\\
&&I_{1,\left(2 \alpha _5+\alpha _6\right) \alpha _4}=-2 (D-1) (3 D-7) (5 D-11) \tilde{r}_0^{4 D}-(D-2) (15 D^3-297 D^2+1157 D
\cr
&&\quad -1259) \tilde{q}^2 \tilde{r}_0^{2 D+6}+\left(25 D^4-449 D^3+2421 D^2-5243 D+4014\right) \tilde{q}^4 \tilde{r}_0^{12}\,,
\nn\\
&&I_{1,\left(2 \alpha _5+\alpha _6\right)^2}=(5 D-13) (D^2-16 D+31) \tilde{q}^2 \tilde{r}_0^6+32 (D-2)^2 \tilde{r}_0^{2 D}\ .
\eea
Adding up \eqref{action1} and \eqref{action2}, we obtain the complete on-shell action of the charged static black hole with corrections from both 4- and 6-derivative terms, namely
\be
I_E(T_0,\Psi_{e,0})=I_0[\phi_{i(0)}]+\epsilon I_{hd1}\left[\phi_{i(0)}+\frac{1}{2}\epsilon\phi_{i(1)}\right]+\epsilon^2I_{hd2}[\phi_{i(0)}]\ .
\ee
Using the free energy $F(T_0,\Psi_{e,0})=T_0 I_E(T_0,\Psi_{e,0})$, we obtain the perturbed entropy and mass
\bea
S&=&-\left(\frac{\partial F(T_0,\Psi_{e,0})}{\partial T_0}\right)\Big|_{\Psi_{e,0}},\quad Q_e=-\left(\frac{\partial F(T_0,\Psi_{e,0})}{\partial \Psi_{e,0}}\right)\Big|_{T_{0}},\nn\\ M&=&F(T_0,\Psi_{e,0})+T_0S+\Psi_{e,0}Q_e\,,
\label{sqm}
\eea
up to and including ${\cal O}(\alpha_i^2)$, even though we have only perturbative solutions up to and including the ${\cal O}(\alpha_i)$'s order.

In order to obtain the perturbed thermodynamic variables with both conserved quantities, namely the mass and charge fixed, we need to perform another redefinition of the integration constants, given by \eqref{parameter redefinition}. The mass and charge take the same form as the original unperturbed solution
\bea
M'=\frac{(D-2)\omega_{D-2}}{16\pi}(r_0^{D-3}+\frac{q^2}{r_0^{D-3}}),\qquad Q_e'=\frac{\sqrt{2(D-2)(D-3)}\omega_{D-2}}{16\pi}q\,.
\label{fix mass and charge cubic}
\eea
We present all the remaining thermodynamic variables up to and including the NNLO in Appendix \ref{charge shift}, due to their lengthy expressions.  In particular, with the explicit entropy result, we can also verify the entropy shift formula \eqref{entropy shift}.

%%%%%%%%%%%%%%%%%%%%%%%%%%%%%%%%%%%%%%%%
\subsection{Poor man's approach by solving second-order perturbative solutions}
%%%%%%%%%%%%%%%%%%%%%%%%%%%%%%%%%%%%%%%

We now show that the results obtained in the previous subsection correctly produce the one that can be directly derived from the ordinary method by solving for the perturbed solutions with fixed mass and charge, up to and including ${\cal O}(\alpha^2_i)$, namely
\bea
&&h=h_0+\Delta h_1+\Delta h_2\,,\qquad   f =f_0+\Delta f_1+\Delta f_2\,,\qquad \psi =\psi_0+\Delta \psi_1+\Delta \psi_2\,,
\label{charge solution cubic}
\eea
where the first and second order deviations from the leading order solution are given in \eqref{Maxwell quadratic} and \eqref{Maxwell cubic} in the Appendix respectively. These perturbations leave the mass and charge fixed, as in \eqref{fix mass and charge cubic}, but the radius of horizon receives following corrections, up to and including ${\cal O}(\alpha_i^2)$
\be
r_h=r_0+\delta_1 r_h +\delta_2 r_h\ .
\label{horizon shift cubic}
\ee
where $\delta_1 r_h$ and $\delta_1 r_h$ are given in \eqref{horizon shift} in the Appendix.
With the explicit NNLO solution, all the remaining thermodynamic quantities such as temperature, entropy and electrostatic potential can be obtained straightforwardly. We find that they match precisely those obtained in the earlier subsection using only the NLO solution, thereby confirming our simpler approach.

\section{Higher-order cases}

In the previous sections, we derived the formalism to obtain thermodynamic contributions up to and including NNLO, using only the NLO perturbative solutions. This construction can be formally generalized to all orders. Consider
\be
\mathcal{L}=\sum_{n=0}^{N}\epsilon^n\mathcal{L}_n\,,\qquad \phi_i=\sum_{i=0}^{N}\epsilon^n\phi_{i(n)}\,.
\ee
After adding the surface term, the Euclidean action is
\bea
I^{(N)}[\phi_i]=\sum_{n=0}^{N}\epsilon^nI_n[\phi_i].
\eea
We should expand all the $I_n[\phi_i]$, $n\in[0,N]$ as the functions of $\phi_{i(0)}$ to the $n$'th order. The formalism requires us to employ perturbative solutions with the temperature $T$, angular velocity $\Omega$, electricity potential $\Psi_e$ fixed, i.e.
\bea
T=T_0+\mathcal{O}(\epsilon^{n+1})\,,\qquad \Omega=\Omega_0+\mathcal{O}(\epsilon^{n+1})\,,\qquad \Psi_e=\Psi_{e,0}+\mathcal{O}(\epsilon^{n+1})\,,\qquad\cdots\,.
\eea
In this case, the EOMs of all orders will give us $N$ constraints
\bea
\frac{\delta I^{(N-1)}[\phi_i]}{\delta\phi_i}\Big|_{\phi_i=\sum_{i=0}^{N-1}\epsilon^n\phi_{i(n)}}=0.
\eea
It's hard to give its general expressions. Instead we give some low-lying examples
\bea
\epsilon^0:\quad &&I_0[\phi_{i(0)}],\cr
\epsilon^1:\quad &&I_1[\phi_{i(0)}],\cr
\epsilon^2:\quad &&I_2[\phi_{i(0)}]+\ft12\int d^Dx
\phi_{i(1)}(x)\frac{\delta I_{1}}{\delta \phi_{i}(x)}\Big|
_{\phi_{i}=\phi_{i(0)}},\cr
\epsilon^3:\quad &&I_3[\phi_{i(0)}]+\ft23\int d^Dx
\phi_{i(1)}(x)\frac{\delta I_{2}}{\delta \phi_{i}(x)}\Big|
_{\phi_{i}=\phi_{i(0)}}+\ft13\int d^Dx
\phi_{i(2)}(x)\frac{\delta I_{1}}{\delta \phi_{i}(x)}\Big|
_{\phi_{i}=\phi_{i(0)}}\cr
&&+\ft16\int d^Dx\int d^Dy
\phi_{i(1)}(x)\phi_{j(1)}(y)\frac{\delta^2 I_{1}}{\delta \phi_{i}(x)\delta \phi_{j}(y)}\Big|
_{\phi_{i}=\phi_{i(0)}},\cr
\epsilon^4:\quad &&I_4[\phi_{i(0)}]+\ft12\int d^Dx
\phi_{i(2)}(x)\frac{\delta I_{2}}{\delta \phi_{i}(x)}\Big|
_{\phi_{i}=\phi_{i(0)}}+\ft34\int d^Dx
\phi_{i(1)}(x)\frac{\delta I_{3}}{\delta \phi_{i}(x)}\Big|
_{\phi_{i}=\phi_{i(0)}}\cr
&&+\ft14\int d^Dx
\phi_{i(3)}(x)\frac{\delta I_{1}}{\delta \phi_{i}(x)}\Big|
_{\phi_{i}=\phi_{i(0)}}\cr
&&+\ft14\int d^Dx\int d^Dy
\phi_{i(1)}(x)\phi_{j(2)}(y)\frac{\delta^2 I_{1}}{\delta \phi_{i}(x)\delta \phi_{j}(y)}\Big|
_{\phi_{i}=\phi_{i(0)}}\cr
&&+\ft14\int d^Dx\int d^Dy
\phi_{i(1)}(x)\phi_{j(1)}(y)\frac{\delta^2 I_{2}}{\delta \phi_{i}(x)\delta \phi_{j}(y)}\Big|
_{\phi_{i}=\phi_{i(0)}}\cr
&&+\frac{1}{24}\int d^Dx\int d^Dy\int d^Dz
\phi_{i(1)}(x)\phi_{j(1)}(y)\phi_{k(1)}(z)\frac{\delta^3 I_{1}}{\delta \phi_{i}(x)\delta \phi_{j}(y)\delta \phi_{k}(z)}\Big|
_{\phi_{i}=\phi_{i(0)}}.
\eea
This gives a clear pattern that at $\epsilon^n$ order, perturbative solutions $\phi_{i(0)}, \phi_{i(1)}, \cdots, \phi_{i(n-1)}$ are needed. Our earlier explicit demonstration worked up to and including the $\epsilon^2$ order.

\section{Conclusion}
In this work we propose a formula which computes the on-shell Euclidean action of asymptotically flat black holes with NNLO contributions, using only the perturbed solution up to and including the NLO contributions. Using this formula, we obtain thermodynamic quantities of black hole solutions in pure gravity and Einstein-Maxwell theories
in $D$-dimensions with 4- and 6-derivative corrections, using only solutions with 4-derivative corrections. We also verified these results by first solving the perturbed field equations with both 4- and 6-derivative corrections and evaluating the on-shell action on the solutions. Our result indicates that a recent work \cite{Aalsma:2022knj} which computes the 4- and 6-derivative corrections to thermodynamic quantities of Myers-Perry black hole
by evaluating the higher derivative action on the leading order solution is incorrect. In fact, one can readily extend our formula to include all order higher derivative corrections. As a byproduct, we also prove that for asymptotically flat black holes, the thermodynamic quantities
depend only on the higher derivative couplings that are invariant under the field redefinition.

As future directions, it should be interesting to loose the asymptotically flat condition and obtain a similar on-shell action formula for asymptotically AdS black holes.  Applying directly the Real-Santos method for AdS black holes were done in several papers
\cite{Melo:2020amq,Bobev:2022bjm,Cassani:2022lrk}. However, intuitively for AdS black holes, the higher derivative contributions to the solution could modify both the effective cosmological constant and the surface terms in the action. Thus a more careful analysis of both contributions in the action seems unavoidable. Since the real world black holes carry spins, we would also like to study higher derivative corrections to thermodynamics of rotating black holes. In this case, some new technique is needed to solve for the NLO solution with only axial symmetry. This problem is circumvented in odd dimensions with all equal angular momenta where the level surfaces become squashed $U(1)$ bundles of ${\mathbb{CP}}$ spaces. For example, the NNLO solution for five-dimensional rotating black hole with equal angular momentum was given in \cite{Ma:2020xwi}. Finally, our results can also be applied to compute the charge to mass ratio of extremal black holes, providing constraints on the low effective theory of quantum gravity at NNLO. In this regard, it would be useful to compare with other approach specific for extremal black holes \cite{Cano:2023dyg}.

\section*{Acknowledgement}

The work is supported in part by the National Natural Science Foundation of China (NSFC) grants No.~12175164, No.~11935009 and No.~11875200.
Y.P. also acknowledges support
by the National Key Research and Development Program
under grant No. 2022YFE0134300.
\appendix

\section{4- and 6-derivative corrections to the static black holes}
In this section, we present the first and second order perturbative solutions.
\subsection{Pure gravity case}
The fully non-linear equations of motion of the pure gravity case  is
\begin{equation}
\mathcal{E}_{ab}=P_{(a}^{\ \ cde}R_{b)cde}-\frac{1}{2}g_{ab}\mathcal{L}%
-2\nabla^{c}\nabla^{d}P_{(a|c|b)d}\,, \label{EOM}%
\end{equation}
where the tensor $P^{abcd}$ is given by
\bea
P_{abcd}&=&P_{abcd}^E+\sum_{J=1}^{3}\alpha_JP_{abcd}^{Q,J}+\sum_{J=1}^{8}\beta_JP_{abcd}^{C,J}\,,
\cr
P_{abcd}^E&=&\frac{1}{2}\left(  g_{ac}g_{bd}-g_{ad}g_{bc}\right)\,,\qquad
P_{abcd}^{Q,1}=\left(  g_{ac}g_{bd}-g_{ad}g_{bc}\right)R\,,\cr P_{abcd}^{Q,2}&=&\frac{1}{2}\left(g_{bd}R_{ac}-g_{bc}R_{ad}-g_{ad}R_{bc}
+g_{ac}R_{bd}\right)\,,\qquad
P_{abcd}^{Q,3}= 2 R_{abcd}\,,
\cr
P_{abcd}^{C,1}&=&\frac{3}{2}\left(  g_{ac}g_{bd}-g_{ad}g_{bc}\right)R^2\,,
\cr
P_{abcd}^{C,2}&=&\frac{1}{2}\left(  g_{ac}g_{bd}-g_{ad}g_{bc}\right)R_{ef}R^{ef}+\frac{1}{2}R\left(g_{bd}R_{ac}-g_{bc}R_{ad}-g_{ad}R_{bc}
+g_{ac}R_{bd}\right)\,,
\cr
P_{abcd}^{C,3}&=&\frac{3}{4}(g_{ac}R_{be}R_d^{\ e}-g_{ad}R_{be}R_c^{\ e}-g_{bc}R_{ae}R_d^{\ e}+g_{bd}R_{ae}R_c^{\ e})\,,
\cr
P_{abcd}^{C,4}&=&\frac{1}{2}(g_{bd}R_{aecf}-g_{bc}R_{aedf}-g_{ad}R_{becf}+g_{ac}R_{bedf})R^{ef}+\frac{1}{2}\left(  R_{ac}R_{bd}-R_{ad}R_{bc}\right)\,,
\cr
P_{abcd}^{C,5}&=&\frac{1}{2}\left(  g_{ac}g_{bd}-g_{ad}g_{bc}\right)R_{efgh}R^{efgh}+2RR_{abcd}\,,
\cr
P_{abcd}^{C,6}&=&\frac{1}{2}\left(R_b^{\ e}R_{aicd}+R_d^{\ e}R_{abce}-R_c^{\ e}R_{abde}-R_a^{\ e}R_{bicd}\right)
\cr
&&+\frac{1}{4}(g_{bd}R_{a}^{\ efg}R_{cefg}-g_{ad}R_{b}^{\ efg}R_{cefg}+g_{ac}R_{b}^{\ efg}R_{defg}-g_{bc}R_{a}^{\ efg}R_{defg})\,,\cr
P_{abcd}^{C,7}&=& 3R_{ab}^{\ \ ef}R_{cdef}\,,\qquad
P_{abcd}^{C,8}=\frac{3}{2}(R_{a\ c}^{\ e\ \ f}R_{bedf}-R_{a\ d}^{\ e\ \ f}R_{becf})\ .
\eea

The deviations from the leading order solution are given by
\bea
\Delta f_1 &=&\Delta h_1=\frac{ (D-4) (D-3) \mu ^2}{r^{2 (D-2)}}\alpha _3\,,
\nn\\
\Delta f_2 &=&\frac{(D-3) (D-1)^2 \mu ^2 }{ 2r^{2 (D-1)}} \Big(8  (D-2)\beta _5+ (3 D-8)\beta _6+12  (D-3)\beta _7\Big)
\cr
&&-\frac{(D-3) \mu ^3}{8 r^{3 D-5}}\Big[8  (D-2) (D-1) (3 D-1)\beta _5+ 4  (D-1) \left(2 D^2-5 D-1\right)\beta _6\cr
&&+4  \left(7 D^3-21 D^2-16 D+50\right)\beta _7 - \left(3 D^2-21 D+38\right)\beta _8
\Big]
\cr
&&-2 \alpha _3 (D-4) (D-3) \mu ^2\Big[\frac{ (D-1)^2 }{r^{2 (D-1)}}\left(4 \alpha _1+3 \alpha _2+8 \alpha _3\right)
\cr
&&-\frac{\mu  }{r^{3 D-5}}\left(  (D-1) (3 D-1)\alpha _1+2  D(D-1)\alpha _2 +\left(4 D^2+3 D-13\right)\alpha _3\right)
\Big]\,,
\nn\\
\Delta h_2 &=&-\frac{(D-3) (D-2) (D-1) \mu ^2}{4 r^{2 (D-1)}} \Big[8 \beta _5-2  (D-5)\beta _6-12  (D-4)\beta _7-3 \beta _8\Big]
\cr
&&+\frac{(D-3)\mu ^3 }{ 8r^{3 D-5}}\Big[8  (D-2) (D-1)^2\beta _5+4  (D-1)^2\beta _6-4  \left(D^3-3 D^2-16 D+38\right)\beta _7
\cr
&&- \left(3 D^2+3 D-26\right)\beta _8
\Big]+2 \alpha _3 (D-4) (D-3) \mu ^2\Big[\frac{(D-1)}{r^{2 (D-1)}}\big(2 \alpha _1- (D-3)
\alpha_2
\cr
&&-2 (2 D-5)\alpha _3 \big)-\frac{\mu }{r^{3 D-5}}\left( (D-1)^2\alpha _1- (5 D-11)\alpha _3\right)
\Big]\ .
\label{cubic perturbed static solution}
\eea

\subsection{Einstein-Maxwell case}
The fully non-linear equations of motion derived from (\ref{Einstein-Maxwell}) are given by
\bea
\mathcal{E}_{ab}&&=P_{(a}^{\ \ cde}R_{b)cde}-\frac{1}{2}g_{ab}\mathcal{L}%
-2\nabla^{c}\nabla^{d}P_{(a|c|b)d} +\mathcal{M}_{(a|c}F_{b)}^{\ \ c},\quad
S^a=-2\nabla_b\mathcal{M}^{ba} ,
\eea
where $\mathcal{M}_{ab}:=\frac{\delta \mathcal{L}}{\delta F^{ab}}$.
In our example, $P_{abcd}$ and $\mathcal{M}_{ab}$ are
\bea
\mathcal{M}_{ab}&=&\mathcal{M}_{ab}^E+\sum_{I=3}^{6}\alpha_I\mathcal{M}_{ab}^{Q,I}
+\beta_7\mathcal{M}_{ab}^{C,1}+\beta_8\mathcal{M}_{ab}^{C,2}+
\gamma_1\mathcal{M}_{ab}^{C,3}+\gamma_2\mathcal{M}_{ab}^{C,4}+
\gamma_3\mathcal{M}_{ab}^{C,5}+\gamma_4\mathcal{M}_{ab}^{C,6}\,,
\nn\\
P_{abcd}&=&P_{abcd}^E+\sum_{I=3}^{6}\alpha_IP_{abcd}^{Q,I}+
\beta_7P_{abcd}^{C,1}+\beta_8P_{abcd}^{C,2}+\gamma_1P_{abcd}^{C,3}+\gamma_2P_{abcd}^{C,4}
+\gamma_3P_{abcd}^{C,5}+\gamma_4P_{abcd}^{C,6}\,,
\eea
in which
\bea
P_{abcd}^E&=&\frac{1}{2}\left(  g_{ac}g_{bd}-g_{ad}g_{bc}\right),\qquad \mathcal{M}_{ab}^E=-\frac{1}{2}F_{ab}\,,\qquad
P_{abcd}^{Q,3}=2R_{abcd},\quad \mathcal{M}_{ab}^{Q,3}=0\,,
\cr
P_{abcd}^{Q,4}&=&F_{ab}F_{cd},\qquad \mathcal{M}_{ab}^{Q,4}=2R_{abcd}F^{cd}\,,\qquad
P_{abcd}^{Q,5}=0,\qquad \mathcal{M}_{ab}^{Q,5}=-2F_{ab}F^2\,,\cr
P_{abcd}^{Q,6}&=& 0,\qquad \mathcal{M}_{ab}^{Q,6}=-2F_a^{\ c}F_b^{\ d}F_{cd}\,,\qquad
P_{abcd}^{C,1}=3R_{ab}^{\ \ ef}R_{cdef},\qquad \mathcal{M}_{ab}^{C,1}=\mathcal{M}_{ab}^{C,2}=0\,,\cr
P_{abcd}^{C,2}&=&\frac{3}{2}(R_{a\ c}^{\ e\ \ f}R_{bedf}-R_{a\ d}^{\ e\ \ f}R_{becf}),\qquad P_{abcd}^{C,3}=\frac{1}{2}(F_a^{\ e}F_b^{\ f}F_{cd}+F_c^{\ e}F_d^{\ f}F_{ab})F_{ef}\,,
\cr
\mathcal{M}_{ab}^{C,3}&=&F^{gh}F_a^{\ e}F_b^{\ f}R_{efgh}-F_e^{\ f}F^{gh}F_b^{\ e}R_{afgh}+F_e^{\ g}F^{ef}F_f^{\ h}R_{abgh}+F_e^{\ f}F^{gh}F_a^{\ e}R_{bfgh}\,,
\cr
P_{abcd}^{C,4}&=&\frac{1}{2}\left(F_{ce}F_{df}-F_{cf}F_{de}\right)F_a^{\ e}F_b^{\ f},\qquad \mathcal{M}_{ab}^{C,4}=2F_e^{\ h}F^{ef}\left(F_a^{\ g}R_{bfgh}-F_b^{\ g}R_{afgh}\right)\,,
\cr
P_{abcd}^{C,5}&=&\frac{1}{2}F_e^{\ f}(F_c^{\ e}R_{abdf}+F_a^{\ e}R_{bfcd}-F_b^{\ e}R_{afcd}-F_d^{\ e}R_{abcf})\,,
\cr
\mathcal{M}_{ab}^{C,5}&=&R_{efgh}(F_a^{\ e}R_b^{\ fgh}-F_b^{\ e}R_a^{\ fgh})\,,
\cr
P_{abcd}^{C,6}&=&F_c^{\ e}F_d^{\ f}R_{abef}+F_a^{\ e}F_b^{\ f}R_{cdef},\qquad \mathcal{M}_{ab}^{C,6}=2F^{ef}R_{ae}^{\ \ gh}R_{bfgh}\ .
\eea

The first order corrections to
the static charged black hole solution \eqref{charge solution} are
\bea
\Delta h_1&=&\frac{2 (D-3) \mu  q^2 }{(D-2) r^{3 D-7}}\left[D \alpha_3+(D-3) (D-2) \alpha_4\right]\cr
&&+\frac{(D-3) q^4 }{(D-2) (3 D-7) r^{2 (2 D-5)}}\big[2 \left(4 D^3-34 D^2+87 D-68\right) \alpha_3-4 (D-3)^2 (D-2) \alpha_4\big]\cr
&&+\frac{(D-3) }{(D-2) r^{2 (D-2)}}\big[(D-4) (D-2) \mu ^2 \alpha_3-4 \left(2 D^2-11 D+16\right)  q^2 \alpha_3\cr
&&-4 (D-3) (D-2) q^2 \alpha_4\big]-\frac{ 8 (D-3)^2 (D-2) q^4}{(3 D-7) r^{4 D-10}}\left(2 \alpha _5+\alpha _6\right)\,,
\nn\\
%%%
\Delta f_1&=&\Delta h_1-\frac{4 (D-3) q^2 f_0}{(D-2) r^{2 (D-2)}}\left[\left(2 D^2-9 D+8\right) \alpha_3+(D-2) (D-1) \alpha_4
\right]\,,
\nn\\
\Delta \psi_1&=&-\frac{2 \sqrt{2} (D-3)^{3/2} q^3 }{\sqrt{D-2} (3 D-7) r^{3 D-7}}\left[(D-2) (7 D-19) \alpha_4-\left(2 D^2-9 D+8\right) \alpha_3\right]\cr
&&+\frac{2 \sqrt{2} (D-3)^{3/2} \sqrt{D-2} \mu  q \alpha_4}{r^{2 (D-2)}}-\frac{ 16 \sqrt{2} (D-3)^{3/2} (D-2)^{3/2} q^3}{(3 D-7) r^{3 D-7}}\left(2 \alpha _5+\alpha _6\right)\ .
\label{Maxwell quadratic}
\eea
The second order corrections to the static charged black hole solution \eqref{charge solution} are
\bea
\Delta h_2&=&-\frac{(D-3)\Big(h_{\alpha_3^2}\alpha_3^2+h_{\alpha_3\alpha_4}\alpha_3\alpha_4
+h_{\alpha_4^2}\alpha_4^2+\sum_{i=7}^{8}h_{\beta_i}\beta_i+\sum_{i=1}^{4}h_{\gamma_i}\gamma_i
\Big)}{8 (D-2)^2 (3 D-7) (3 D-5) (5 D-11) r^{6 D-2}}\cr
&&+\frac{16 \left(2 \alpha _5+\alpha _6\right)}{(3 D-7) (5 D-11) r^{6 D-8}}\Big(h_{\alpha_3}\alpha_3+h_{\alpha_4}\alpha_4+h_{2\alpha_5+\alpha_6}(2\alpha_5+\alpha_6)
\Big)\,,
\nn\\
\Delta f_2&=&\frac{(D-3)\Big(f_{\alpha_3^2}\alpha_3^2+f_{\alpha_3\alpha_4}\alpha_3\alpha_4
+f_{\alpha_4^2}\alpha_4^2+\sum_{i=7}^{8}f_{\beta_i}\beta_i+\sum_{i=1}^{4}f_{\gamma_i}\gamma_i
\Big)}{8 (D-2)^2 (3 D-7) (3 D-5) (5 D-11) r^{6 D-2}}\cr
&&+\frac{16 \left(2 \alpha _5+\alpha _6\right) (D-3)^2 q^4 r^{8-6 D}}{(3 D-7) (5 D-11)}\Big(f_{\alpha_3}\alpha_3+f_{\alpha_4}\alpha_4+f_{2\alpha_5+\alpha_6}(2\alpha_5+\alpha_6)
\Big)\,,
\label{Maxwell cubic}
\\
%%%
\Delta \psi_2&=&-\frac{(D-3)^{3/2} q\Big(\psi_{\alpha_3^2}\alpha_3^2+\psi_{\alpha_3\alpha_4}\alpha_3\alpha_4
+\psi_{\alpha_4^2}\alpha_4^2+\sum_{i=7}^{8}\psi_{\beta_i}\beta_i+
\sum_{i=1}^{4}\psi_{\gamma_i}\gamma_i
\Big)}{4 \sqrt{2} (D-2)^{3/2} (3 D-7) (3 D-5) (5 D-11) r^{5 D-5}}\cr
&&+\frac{16 \sqrt{2} \left(2 \alpha _5+\alpha _6\right) (D-3)^{5/2} \sqrt{D-2} q^3 r^{8-5 D}}{(3 D-7) (5 D-11)}(\psi_{\alpha_3}\alpha_3+\psi_{\alpha_4}\alpha_4+\psi_{2\alpha_5+\alpha_6}
(2\alpha_5+\alpha_6)
\Big)\,,\nn
\eea
where
\bea
&&h_{\alpha_3^2}=-16\Big(
4 (D-2)^2 (2 D-5) (3 D-5) \left(3 D^4-6 D^3+22 D^2-193 D+308\right) q^6 r^{12}\cr
&&\ \ -(D-4) (D-2)^2 (3 D-7) (3 D-5) (5 D-11)\mu ^2 r^{3 D} \big(2 (D-1) (2 D-5)  r^D\cr
&&+(11-5 D) \mu  r^3\big)-(5 D-11) q^4 r^{D+6} \big(4 (D-2) (3 D-7)(2 D^5-115 D^4+879 D^3\cr
&&\ \ -2634 D^2+3522 D-1768)  r^D +(682 D^6-8533 D^5+44205 D^4-121567 D^3\cr
&&\ \ +187533 D^2-154184 D+52896) \mu  r^3\big) +2 (D-2) (3 D-7) (5 D-11) q^2 r^{2 D}\cr
&&\ \ \times \big[8 (D-3) (D-1) (3 D-5) \left(2 D^2-9 D+12\right)  r^{2 D} -4(8 D^5-55 D^4+117 D^3\cr
&&\ \ -33 D^2-165 D+148)  \mu  r^{D+3}+(D-2) (12 D^4-67 D^3 +133 D^2-128 D+48) \mu ^2 r^6\big]
\Big)\,,
\nn\\
%%%%
&&h_{\alpha_3\alpha_4}=32 (D-2) q^2\Big(2 (3 D-5) (61 D^6-1035 D^5+6932 D^4-23880 D^3+45061 D^2\cr
&&\ \ -44407 D+17912) q^4 r^{12}-(5 D-11) q^2 r^{D+6} \big[2 (3 D-7) (55 D^5-638 D^4+3015 D^3\cr
&&\ \  -7184 D^2 +8560 D-4052)  r^D+(-69 D^6+756 D^5-3653 D^4+10133 D^3-17026 D^2\cr
&&\ \ +16159 D -6604) \mu  r^3\big]-2 (D-2) (3 D-7) (5 D-11) r^{2 D}\big[4 (D-1) (3 D-5) \cr
&&\ \ \times \left(D^2-5 D+7\right)  r^{2 D} -2 \left(13 D^4-105 D^3+328 D^2-459 D+239\right)  \mu  r^{D+3}\cr
&&\ \ +(D-2) \left(6 D^3-24 D^2+33 D-13\right) \mu ^2 r^6\big]
\Big)\,,
\nn\\
&&h_{\alpha_4^2}=16 (D-3)^2 (D-2)^2 q^2 r^3\Big(4 (3 D-5) \left(16 D^3-83 D^2+118 D-27\right) q^4 r^9\cr
&&\ \ +8 (D-2) (3 D-7) (5 D-11) \mu  r^{2 D} \left(4 (2 D-3)  r^D+(4-3 D) \mu  r^3\right)-(3 D-7)\cr
 &&\ \ \times (5 D-11) q^2 r^{D+3} \left(4 \left(76 D^2-317 D+317\right)  r^D+\left(-121 D^2+508 D-507\right) \mu  r^3\right)
\Big)\,,
\nn\\
&&h_{\beta_{7}}=4 (D-2) (3 D-7)\Big(4 (3 D-5) (28 D^5-360 D^4+1877 D^3-4913 D^2+6400 D\cr
&&\ \ -3296) q^6 r^{12} -(D-2) (3 D-5) (5 D-11) \mu ^2 r^{3 D} \big(6 (D-4) (D-2) (D-1)  r^D\cr
&&\ \ -\left(D^3-3 D^2-16 D+38\right) \mu  r^3\big) +6 (D-2) (5 D-11) \mu  q^2 r^{2 D+3} \big[16 (D-2)^2\cr
&&\ \ \times \left(2 D^2-12 D+17\right)  r^D+(-9 D^4+86 D^3-283 D^2 +390 D-196) \mu  r^3\big]\cr
&&\ \ -6 (5 D-11) q^4 r^{D+6} \big[2 (3 D-5) \left(12 D^4-132 D^3+541 D^2-982 D+668\right) r^D\cr
&&\ \ +\left(-20 D^5+256 D^4-1271 D^3+3099 D^2-3742 D+1796\right) \mu  r^3\big]
\Big)\,,
\nn\\
&&h_{\beta_{8}}=-(D-2)^2 (3 D-7)\Big(4 (3 D-5) \left(69 D^3-471 D^2+1052 D-770\right) q^6 r^{12}\cr
&&\ \ +(3 D-5) (5 D-11) \mu ^2 r^{3 D} \left(\left(-3 D^2-3 D+26\right) \mu  r^3+6 (D-2) (D-1)  r^D\right)\cr
&&\ \ +6 (5 D-11) q^4 r^{D+6} \big(2 (3 D-5) \left(7 D^2-36 D+46\right)  r^D+(-49 D^3+295 D^2  \cr
&&\ \ -570 D+358) \mu  r^3\big)-6 (5 D-11) \mu  q^2 r^{2 D+3} \big(4 (D-2) \left(7 D^2-30 D+31\right)  r^D\cr
&&\ \  -\left(21 D^3-108 D^2+167 D-76\right) \mu  r^3\big)\Big)\,,
\nn\\
&&h_{\gamma_{1}}=2h_{\gamma_{2}}=-48 (D-3)^2 (D-2)^3 (3 D-7) (3 D-5) q^4 r^6\cr
&&\ \ \times \left((5 D-11) r^D \left(2  r^D-\mu  r^3\right)+2 (D-3) q^2 r^6\right)\,,
\nn\\
&&h_{\gamma_{3}}=16 (D-3) (D-2)^2 (3 D-7) q^2 r^3\Big(4 (3 D-5) \left(D^3-10 D^2+26 D-19\right) q^4 r^9\cr
&&\ \ +(D-2) (5 D-11) \mu  r^{2 D} \left(2 \left(4 D^2-17 D+19\right)  r^D+\left(-3 D^2+10 D-11\right) \mu  r^3\right)\cr
&&\ \ -2 (5 D-11) q^2 r^{D+3} \big(2 (3 D-5) \left(3 D^2-16 D+22\right)  r^D\cr
&&\ \ -\left(7 D^3-45 D^2+102 D-78\right) \mu  r^3\big)\Big)\,,
\nn\\
&&h_{\gamma_{4}}=16 (D-3)^2 (D-2)^2 (3 D-7) q^2 r^3\Big(2 (D-1) (2 D-5) (3 D-5) q^4 r^9\cr
&&\ \ +(D-2) (5 D-11) \mu  r^{2 D} \left(4 (2 D-3)  r^D+(4-3 D) \mu  r^3\right)\cr
&&\ \ -(5 D-11) q^2 r^{D+3} \left(\left(-14 D^2+59 D-59\right) \mu  r^3+6 (2 D-5) (3 D-5) r^D\right)
\Big)\,,
\nn\\
&&h_{\alpha_3}=(D-3)^2 q^4\Big(
(5 D-11) r^D \big[2 \left(12 D^3-91 D^2+231 D-196\right) r^D -(8 D^3-57 D^2\cr
&&\ \ +139 D-116) \mu  r^3\big]-2 \left(18 D^4-190 D^3+723 D^2-1181 D+700\right) q^2 r^6
\Big)\,,
\nn\\
&&h_{\alpha_4}=2 (D-3)^3 (D-2) q^4 \big[\left(19 D^2-104 D+139\right) q^2 r^6+3 \left(15 D^2-68 D+77\right)\cr
&&\ \ \times r^D \left(2 r^D-\mu  r^3\right)\big],\cr
&&h_{2\alpha_5+\alpha_6}=16 (3 D-7) (D-3)^3 (D-2)^2 q^6 r^6\,,
\nn\\
&&f_{\alpha_3^2}=16\Big(4 (3 D-5) (176 D^7-2120 D^6+10060 D^5-22040 D^4+14339 D^3+28599 D^2\cr
&&\ \ -57756 D +30272) q^6 r^{12}-(D-4) (D-2)^2 (3 D-7) (3 D-5) (5 D-11) \mu ^2 r^{3 D} \big[\cr
&&\ \ \left(-4 D^2-3 D+13\right) \mu  r^3  +8 (D-1)^2  r^D\big]+(5 D-11) q^4 r^{D+6} \big[16 (D-2) (3 D-7)\cr
&&\ \ \times (22 D^5-185 D^4+575 D^3 -745 D^2+253 D+132)  r^D+(5-3 D) (296 D^6-3206 D^5
\cr
&&\ \ +13843 D^4-29036 D^3 +27115 D^2-3368 D-7264) \mu  r^3\big]+2 (D-2) (3 D-7)\cr
&&\ \ \times (5 D-11) q^2 r^{2 D}\big[16 (D-1)^2 (3 D-5) (2 D^2-11 D+16)  r^{2 D}-4 (3 D-5)
(12 D^4\cr
&&\ \ -83 D^3+204 D^2-193 D+52)  \mu  r^{D+3}+(64 D^5-587 D^4+2235 D^3-4276 D^2\cr
&&\ \ +4010 D-1448) \mu ^2 r^6\big]\Big)\,,
\nn\\
&&f_{\alpha_3\alpha_4}=32 (D-3) (D-2) q^2\Big(4 (D-2) (3 D-5)(412 D^4-3496 D^3+11017 D^2\cr
&&\ \ -15262 D+7828) q^4 r^{12} +(5 D-11) q^2 r^{D+6} \big[8 (D-2) (3 D-7) (67 D^3-397 D^2\cr
&&\ \ +784 D-513)  r^D +(5-3 D) \left(493 D^4-4056 D^3+12440 D^2-16853 D+8508\right) \mu  r^3\big]\cr
&&\ \ +(D-2) (3 D-7) (5 D-11) r^{2 D} \big[\left(113 D^3-528 D^2+801 D-392\right) \mu ^2 r^6
\cr
&&\ \ +16 (D-1)^2 (3 D-5) r^{2 D} -20 (3 D-5) \left(3 D^2-9 D+7\right)  \mu  r^{D+3}\big]
\Big)\,,
\nn\\
&&f_{\alpha_4^2}=16 (D-3) (D-2)^2 q^2 r^3\Big(32 (3 D-5)(138 D^4-1196 D^3+3865 D^2-5516 D\cr
&&\ \ +2931) q^4 r^9 -32 (D-2) (2 D-3) (3 D-7) (3 D-5) (5 D-11) \mu  r^{3 D}-(3 D-7)
 \cr
&&\ \ \times (3 D-5) (5 D-11) (383 D^2-1580 D+1605) \mu  q^2 r^{D+6}+8 (D-2) (3 D-7)\cr
&&\ \ \times (5 D-11) r^{2 D+3} \big[4 \left(38 D^2-157 D+156\right) q^2 +\left(19 D^2-53 D+36\right) \mu ^2\big]
\Big)\,,
\nn\\
&&f_{\beta_{7}}=4 (D-2) (3 D-7)\Big(8 (3 D-5) (256 D^5-3024 D^4+14186 D^3-33020 D^2\cr
&&\ \ +38107 D-17426) q^6 r^{12}+(D-2) (3 D-5) (5 D-11) \mu ^2 r^{3 D} \big((-7 D^3+21 D^2\cr
&&\ \ +16 D-50) \mu  r^3+12 (D-3) (D-1)^2  r^D\big) +6 (3 D-5) (5 D-11) q^4 r^{D+6}\cr
&&\ \ \times \big[32 (D-3) (D-2) \left(3 D^2-15 D+19\right) r^D
+(-100 D^4+948 D^3-3343 D^2\cr
&&\ \ +5198 D-3004) \mu  r^3\big]-12 (D-2) (5 D-11) \mu  q^2 r^{2 D+3}\big[(-38 D^4+306 D^3\cr
&&\ \ -893 D^2+1120 D-509) \mu  r^3+16 (D-3) (D-2)^2 (3 D-5) r^D\big]
\Big)\,,
\nn\\
&&f_{\beta_{8}}=(D-2)^2 (3 D-7) r^3\Big(8 (3 D-5) \left(12 D^3-66 D^2+97 D-22\right) q^6 r^9+(3 D-5)\cr
&&\ \ \times (5 D-11) (3 D^2-21 D+38) \mu ^3 r^{3 D}+6 (3 D-5) (5 D-11) q^4 r^{D+3} \big((-5 D^2\cr
&&\ \ +16 D-2) \mu  r^3+8 (D-3) (D-2) r^D\big) +12 (5 D-11) \mu  q^2 r^{2 D} \big((D^3+D^2\cr
&&\ \ -23 D+31) \mu  r^3-2 (D-3) (D-2) (3 D-5)  r^D\big)\Big)\,,
\nn\\
&&f_{\gamma_{1}}=2f_{\gamma_{2}}=48 (D-3) (D-2)^3 (3 D-7) (3 D-5) q^4 r^6 \cr
&&\ \ \times\big[32 (D-2)^2 q^2 r^6+(5 D-11) r^D \left(8 (D-2) r^D+(13-7 D) \mu  r^3\right)\big]\,,
\nn\\
&&f_{\gamma_{3}}=16 (D-3) (D-2)^2 (3 D-7) q^2 r^3\Big(4 (3 D-5) (44 D^3-284 D^2+603 D\cr
&&\ \ -421) q^4 r^9 +2 (3 D-5) (5 D-11) q^2 r^{D+3} \big(\left(-23 D^2+98 D-102\right) \mu  r^3+2 (D-2)\cr
&&\ \ \times (12 D-31)  r^D\big) -(D-2) (5 D-11) \mu  r^{2 D} \big(\left(-19 D^2+58 D-43\right) \mu  r^3\cr
&&\ \ +2 (3 D-5) (4 D-7) r^D\big)\Big)\,,
\nn\\
&&f_{\gamma_{4}}=16 (D-3) (D-2)^2 (3 D-7) q^2 r^3\Big(4 (2 D-5) (3 D-5) \left(22 D^2-85 D+81\right) q^4 r^9\cr
&&\ \ +(3 D-5) (5 D-11) q^2 r^{D+3} \left(\left(-46 D^2+191 D-195\right) \mu  r^3+24 (D-2) (2 D-5) r^D\right)\cr
&&\ \ -(D-2) (5 D-11) \mu  r^{2 D} \left(\left(-19 D^2+53 D-36\right) \mu  r^3+4 (2 D-3) (3 D-5) r^D\right)
\Big)\,,
\nn\\
&&f_{\alpha_3}=(5 D-11) r^D \big[16 \left(6 D^3-41 D^2+93 D-70\right) r^D-(80 D^3-531 D^2+1165 D\cr
&&\ \ -844) \mu  r^3\big] +2 \left(172 D^4-1488 D^3+4789 D^2-6787 D+3568\right) q^2 r^6\,,
\nn\\
&&f_{\alpha_4}=6 (D-2) \big[2 (D-2)^2 (49 D-115) q^2 r^6+(3 D-7) (5 D-11) r^D\cr
&&\ \ \times \left(8 (D-2) r^D-(7 D-13) \mu  r^3\right)\big]\,,
\nn\\
&&f_{2\alpha_5+\alpha_6}=16 (D-3) (D-2)^2 (3 D-7) q^2 r^6\,,
\nn\\
&&\psi_{\alpha_3^2}=16 (3 D-7)\Big((3 D-5) \left(16 D^5-102 D^4+129 D^3+457 D^2-1400 D+1056\right) q^4 r^6 \cr
&&\ \ -(D-4) (D-2)^2 (D-1) (2 D+1) (5 D-11) \mu ^2 r^{2 D} +4 (D-2) (5 D-11) q^2 r^D \cr
&&\ \ \times\big[2 (D-1) \left(2 D^3-11 D^2+15 D+4\right) r^D+\left(-4 D^4+31 D^3-94 D^2+122 D-56\right) \mu  r^3\big]\Big)\,,
\nn\\
&&\psi_{\alpha_3\alpha_4}=32 (D-2)\Big((3 D-5)(191 D^5-2377 D^4+11616 D^3-27963 D^2+33237 D\cr
&&\ \ -15628) q^4 r^6 +2 (D-4) (D-2)^2 (2 D-3) (3 D-7) (5 D-11) \mu ^2 r^{2 D}
-2 (D-2)(3 D-7)\cr
&&\ \ \times (5 D-11) q^2 r^D \big[2 \left(7 D^3-52 D^2+128 D-97\right) r^D+\left(2 D^3-13 D^2+11 D+8\right) \mu  r^3\big]
\Big)\,,
\nn\\
&&\psi_{\alpha_4^2}=-16 (D-3) (D-2)^2\Big((3 D-5) \left(173 D^3-1517 D^2+4331 D-4043\right) q^4 r^6\cr
&&\ \ +8 (D-2)^2 (3 D-7) (5 D-11) \mu ^2 r^{2 D}+4 (3 D-7) (5 D-11) q^2 r^D\cr
 &&\ \ \times\big[\left(-17 D^2+73 D-74\right) \mu  r^3+4 (D-2) (2 D-3) r^D\big]
\Big)\,,
\nn\\
&&\psi_{\beta_{7}}=12 (D-2) (3 D-7)\Big(2 (3 D-5) \left(36 D^4-348 D^3+1251 D^2-1978 D+1156\right) q^4 r^6\cr
&&\ \ -4 (D-2)^2 (5 D-11) \left(4 D^2-16 D+13\right) \mu  q^2 r^{D+3}+(D-2) (D-1) (5 D-11)
\cr
&&\ \ \times\left(D^2-2 D-2\right) \mu ^2 r^{2 D}\Big)\,,
\nn\\
&&\psi_{\beta_{8}}=-3 (D-2)^2 (3 D-7)\Big(2 (3 D-5) \left(3 D^2-16 D+22\right) q^4 r^6\cr
&&\ \ -4 (D-2)^2 (5 D-11) \mu  q^2 r^{D+3} +(D-2) (D-1) (5 D-11) \mu ^2 r^{2 D}
\Big)\,,
\nn\\
&&\psi_{\gamma_{1}}=2\psi_{\gamma_{2}}=-16 (D-3) (D-2)^3 (3 D-5) (3 D-7) q^2 r^3\cr
 &&\ \ \times\left((23 D-65) q^2 r^3-2 (5 D-11) \mu  r^D\right)\,,
 \nn\\
&&\psi_{\gamma_{3}}=16 (D-3) (D-2)^2 (3 D-7)\Big(4 (D-2)^2 (5 D-11) \mu  q^2 r^{D+3}\cr
&&\ \ +2 (D-3) (D+2) (3 D-5) q^4 r^6-(D-2) (D-1) (5 D-11) \mu ^2 r^{2 D}
\Big)\,,
\nn\\
&&\psi_{\gamma_{4}}=16 (D-3) (D-2)^2 (3 D-7)\Big(2 (5 D-11) \left(2 D^2-10 D+11\right) \mu  q^2 r^{D+3}\cr
&&\ \ -(D-2)^2 (5 D-11) \mu ^2 r^{2 D}+(D+5) (2 D-5) (3 D-5) q^4 r^6
\Big)\,,
\nn\\
&&\psi_{\alpha_3}=(3 D-7) \left(-16 D^2+69 D-68\right) q^2 r^3\,,
\nn\\
&&\psi_{\alpha_4}=4 (D-2) \left(\left(41 D^2-211 D+269\right) q^2 r^3-(3 D-7) (5 D-11) \mu  r^D\right)\,,
\nn\\
&&\psi_{2\alpha_5+\alpha_6}=48 (3 D-7) (D-2)^2 q^2 r^3\ .
\eea

\section{4- and 6-derivative corrections to the thermodynamics}\label{charge shift}

To compute the thermodynamic quantities, one needs to first figure out the location of the horizon. We find that for the modified static charged black hole solution \eqref{charge solution cubic}, the horizon \eqref{horizon shift cubic} is located at
\bea
\delta_1 r_h&=&-\frac{1}{(D-2) (3 D-7) ( r_0^{2 D}-q^2 r_0^6)r_0^{2 D+1}}\Big[
\Big((D-4) (D-2) (3 D-7)  r_0^{4 D}\cr
&&-2 (D-3) (3 D-8) (3 D-7)  q^2 r_0^{2 D+6}+(D-3) (D-2) (11 D-32) q^4 r_0^{12}
\Big)\alpha_3\cr
&&+2 (D-3) (D-2) q^2 r_0^6\Big((D-1) q^2 r_0^6-(3 D-7)  r_0^{2 D}
\Big)\alpha_4\cr
&&-8  (D-3) (D-2)^2 q^4 r_0^{12}(2 \alpha _5+\alpha _6)\Big]\,,
\nn\\
\delta_2 r_h&=&\frac{r_{h,\beta_7}\beta_{7}+r_{h,\beta_8}\beta_{8}+r_{h,2 \gamma_{1}+\gamma_{2}}(2 \gamma_{1}+\gamma_{2})+r_{h,\gamma_{3}}\gamma_{3}+r_{h,\gamma_{4}}\gamma_{4}}{8 (D-2) (3 D-5) (5 D-11)r_0^{4 D+3} \left( r_0^{2 D}-q^2 r_0^6\right)}\label{horizon shift}\\
&&+\frac{r_{h,\alpha_3^2}\alpha_3^2+r_{h,\alpha_3\alpha_4}\alpha_3\alpha_4+r_{h,\alpha_4^2}\alpha_4^2}{2 (D-2)^2 (3 D-7)^2 (3 D-5) (5 D-11)r_0^{4 D+3} \left(r_0^{2 D}-q^2 r_0^6\right){}^3}\cr
&&-\frac{8 \left(2 \alpha _5+\alpha _6\right) (D-3) q^4 \big(
r_{h,\alpha_3(2 \alpha _5+\alpha _6)}\alpha_3+r_{h,\alpha_4(2 \alpha _5+\alpha _6)}\alpha_4+r_{h,(2 \alpha _5+\alpha _6)^2}(2 \alpha _5+\alpha _6)
\big)}{(3 D-7)^2 (5 D-11)r_0^{4 D-9} \left(r_0^{2 D}-q^2 r_0^6\right)^3}.\nn
\eea
where
\bea
&&r_{h,\beta_7}=4\Big[3 (D-3) (D-2) (5 D-11) (37 D^3-267 D^2+628 D-478)  q^2 r_0^{4 D+6}\cr
&&\ \ -(D-2) (3 D-5) (5 D-11) (5 D^3-39 D^2+100 D-86)  r_0^{6 D}\cr
&&\ \ -3 (D-3)^2 (5 D-11) \left(79 D^3-530 D^2+1166 D-836\right)  q^4 r_0^{2 (D+6)}\cr
&&\ \ +(D-3)^2 \left(681 D^4-6213 D^3+21024 D^2-31226 D+17156\right) q^6 r_0^{18}
\Big]\,,
\nn\\
&&r_{h,\beta_8}=(D-2)\Big[3 (D-3) (5 D-11) \left(11 D^2-53 D+58\right)  q^2 r_0^{4 D+6}\cr
&&\ \ -(3 D-5) (5 D-11) (3 D^2-21 D+38)  r_0^{6 D}-3 (D-3)^2 (5 D-11) (11 D-18)  q^4 r_0^{2 (D+6)}\cr
&&\ \ +3 (D-3)^2 (D-2) (19 D-33) q^6 r_0^{18}
\Big]\,,
\nn\\
&&r_{h,2 \gamma_{1}+\gamma_{2}}=24 (D-3)^2 (D-2)^2 (3 D-5) q^4 r_0^{12}  \left[(3 D-5) q^2 r_0^6-(5 D-11)  r_0^{2 D}\right]\,,
\nn\\
&&r_{h,\gamma_{3}}=16 (D-3)^2 (D-2) q^2 r_0^6\Big[(D-2) \left(67 D^2-296 D+309\right) q^4 r_0^{12}\cr
&&\ \ +(D-2) (5 D-11) (5 D-9)  r_0^{4 D}-4 (5 D-11) \left(5 D^2-21 D+21\right)  q^2 r_0^{2 D+6}
\Big]\,,
\nn\\
&&r_{h,\gamma_{4}}=16 (D-3)^2 (D-2) q^2 r_0^6\Big[(67 D^3-428 D^2+894 D-611) q^4 r_0^{12}\cr
&&\ \ +(D-2) (5 D-11) (5 D-8)  r_0^{4 D}-(5 D-11) (20 D^2-83 D+83)  q^2 r_0^{2 D+6}
\Big]\,,
\nn\\
&&r_{h,\alpha_3^2}=\Big[(D-3)^2 (1911 D^7-25094 D^6+134863 D^5-371568 D^4+521296 D^3\cr
&&\ \ -261316 D^2 -147024 D+159808) q^{10} r_0^{30}+(D-4) (D-2)^2 (3 D-7)^2 (3 D-5)\cr
&&\ \ \times (5 D-11) (13 D^2-58 D +60)  r_0^{10 D}-2 (D-3) (3 D-7) (5 D-11) (15 D^6\cr
&&\ \ -401 D^5+4164 D^4-20110 D^3+48902 D^2 -58240 D+27008)  q^4 r_0^{6 (D+2)}\cr
&&-(D-3)^2 (8457 D^7-120940 D^6+736857 D^5-2464858 D^4  +4851996 D^3\cr
&&\ \ -5561740 D^2+3382128 D-817344)q^8 r_0^{2 D+24}+2 (D-3) (3 D-7) (1641 D^7\cr
&&\ \ -25820 D^6+175717 D^5-667408 D^4+1519476 D^3-2061378 D^2+1533904 D\cr
&&\ \ -480128)  q^6 r_0^{4 D+18}-(D-4) (D-2) (3 D-7)^2 (5 D-11) (113 D^4-873 D^3\cr
&&\ \ +2366 D^2 -2600 D+936)  q^2 r_0^{8 D+6}\Big]\,,
\nn\\
&&r_{h,\alpha_3\alpha_4}=4 (D-3) (D-2) q^2 r_0^6\Big[(D-3) (D-2) (3426 D^5-38047 D^4+167641 D^3\cr
&&\ \ -366341 D^2+397013 D-170668) q^8 r_0^{24}+(D-2) (3 D-7)^2 (5 D-11) (38 D^3\cr
&& -275 D^2 +639 D-476)  r_0^{8 D}-2 (D-3) (6486 D^6-85396 D^5+465933 D^4-1348471 D^3\cr
&&\ \ +2183197 D^2 -1874593 D+666804)  q^6 r_0^{2 D+18}+2 (3 D-7) (2916 D^6-40397 D^5\cr
&&\ \ +231264 D^4-700171 D^3 +1182150 D^2-1055070 D+388768)  q^4 r_0^{4 D+12}-2 (3 D-7)\cr
&&\ \  \times (5 D-11) (330 D^5-3854 D^4 +17805 D^3-40668 D^2+45907 D-20476)
q^2 r_0^{6 D+6}\Big]\,,
\nn\\
&&r_{h,\alpha_4^2}=4 (D-3)^2 (D-2)^2 q^2 r_0^6\Big[(4926 D^5-54227 D^4+237548 D^3
-517410 D^2\cr
&&\ \ +560086 D-240907) q^8 r_0^{24}+8 (D-2) (3 D-7)^2 (5 D-11) (5 D-8)  r_0^{8 D}\cr
&&\ \ -(17322 D^5-190393 D^4 +833252 D^3-1814294 D^2+1964402 D-845617)  q^6 r_0^{2 D+18}\cr
&&\ \ +(3 D-7) (3 D-5) (2414 D^3-16737 D^2+38614 D-29643)  q^4 r_0^{4 D+12}\cr
&&\ \ -(3 D-7)^2 (5 D-11) \left(250 D^2-989 D+953\right)  q^2 r_0^{6 D+6}
\Big]\,,
\nn\\
&&r_{h,\alpha_3(2 \alpha _5+\alpha _6)}=\Big[(D-3) (767 D^4-6959 D^3+23676 D^2-35806 D+20312) q^4 r_0^{2 (D+6)}\cr
&&\ \ -(D-2) (3 D-7) \left(307 D^3-2393 D^2+6192 D-5316\right) q^2 r_0^{4 D+6}\cr
&&\ \ +(D-3) (D-2) \left(-211 D^3+1463 D^2-3394 D+2636\right) q^6 r_0^{18}\cr
&&\ \ +(3 D-7) (5 D-11) (27 D^3-208 D^2+534 D-456) r_0^{6 D}
\Big]\,,
\nn\\
&&r_{h,\alpha_4(2 \alpha _5+\alpha _6)}=-2 (D-3) (D-2)\Big[
(3 D-7) \left(212 D^2-927 D+1009\right) q^2 r_0^{4 D+6}\cr
&&\ \ +\left(146 D^3-937 D^2+1986 D-1387\right) q^6 r_0^{18}-6 (3 D-7)^2 (5 D-11) r_0^{6 D}\cr
&&\ \ -4 (133 D^3-878 D^2+1926 D-1403) q^4 r_0^{2 (D+6)}
\Big]\,,
\nn\\
&&r_{h,(2 \alpha _5+\alpha _6)^2}=4 (D-3) (D-2)^2 q^2 r_0^6 \Big[-\left(109 D^2-505 D+586\right) q^2 r_0^{2 D+6}\cr
&&\ \ +\left(47 D^2-221 D+260\right) q^4 r_0^{12}+8 (3 D-7)^2 r_0^{4 D}\Big]\ .
\eea
Sticking with the same parameters that leads to the mass and electric charge \eqref{fix mass and charge cubic}, the Hawking  temperature takes the form
\bea
T'&=&\frac{(D-3) }{4 \pi  r_0^{2 D-5}}\left( r_0^{2 (D-3)}-q^2\right)+\Delta T_1+\Delta T_2\,,
\nn\\
\Delta T_1&=&-\frac{(D-3) r_0^{-4 D-3}}{4 \pi  (D-2) (3 D-7) ( r_0^{2 D}-q^2 r_0^6)}\Big[
\Big((D-3) (D-2)(7 D^2-36 D+48) q^6 r_0^{18}\cr
&&+(D-4) (D-2)^2 (3 D-7)  r_0^{6 D}-(D-3) (15 D^3-114 D^2+300 D-272)  q^4 r_0^{2 (D+6)}\cr
&&-(D-4) (D-2)^2 (3 D-7)  q^2 r_0^{4 D+6}
\Big)\alpha_3+2 (D-3) (D-2) q^2 r_0^6\Big((D-2)\cr
&& \times (5 D-13) q^4 r_0^{12}+(D-2) (3 D-7)  r_0^{4 D}-4 (3 D^2-15 D+19)  q^2 r_0^{2 D+6}
\Big)\alpha_4\cr
&&-8  (D-3) (D-2)^2 q^4 r_0^{12} ((2-D) q^2 r_0^6+(3 D-8) r_0^{2 D})(2 \alpha _5+\alpha _6)
\Big]\,,
\nn\\
%%%
\Delta T_2&=&\frac{(D-3) \big(T_{2,\beta_7}\beta_{7}+T_{2,\beta_8}\beta_{8}+T_{2,2\gamma_{1}+\gamma_{2}}(2\gamma_{1}+\gamma_{2})+T_{2,\gamma_3}\gamma_{3}+T_{2,\gamma_4}\gamma_{4}
\big)}{32 \pi  (D-2) (3 D-5) (5 D-11)r_0^{6 D+5} \left(r_0^{2 D}-q^2 r_0^6\right)}\cr
&&+\frac{(D-3) \big(T_{2,\alpha_3^2}\alpha_3^2+T_{2,\alpha_3\alpha_4}\alpha_3\alpha_4+T_{2,\alpha_4^2}\alpha_4^2
\big)}{8 \pi  (D-2)^2 (3 D-7)^2 (3 D-5) (5 D-11) r_0^{6 D+5}\left( r_0^{2 D}-q^2 r_0^6\right){}^3}\cr
&&+\frac{2  \left(2 \alpha _5+\alpha _6\right) (D-3)^2 q^4 r_0^7}{\pi  (3 D-7)^2 (5 D-11) \left(r_0^{2 D}-q^2 r_0^6\right){}^3}\Big[T_{2,\alpha _3 (2 \alpha _5+\alpha _6)}\alpha _3+2 (D-3) (D-2)\cr
&&\times r_0^{-6 D}T_{2,\alpha _4 (2 \alpha _5+\alpha _6)}\alpha _4-4 (D-3) (D-2)^2 q^2 r_0^6T_{2, (2 \alpha _5+\alpha _6)^2} \left(2 \alpha _5+\alpha _6\right)\Big]\,,
\eea
where
\bea
&&T_{2,\beta_7}=4\Big[3 (D-3)^2 (D-2) (129 D^4-1239 D^3+4410 D^2-6886 D+3964) q^8 r_0^{24}\cr
&&\ \ -(D-6) (D-2) (3 D-5) (5 D-11) (D^3-9 D^2+26 D-22) r_0^{8 D}\cr
&&\ \ +6 (D-3) (D-2) (5 D-11) (14 D^4-130 D^3+443 D^2-662 D+372)  q^4 r_0^{4 (D+3)}\cr
&&\ \ -2 (D-3)^2(405 D^5-4623 D^4+20988 D^3-47410 D^2+53272 D-23784)  q^6 r_0^{2 (D+9)}\cr
&&\ \ -6 (D-2)^2 (5 D-11) \left(D^4-2 D^3-31 D^2+122 D-110\right)  q^2 r_0^{6 D+6}
\Big]\,,
\nn\\
&&T_{2,\beta_8}=(D-2)\Big[9 (D-3)^2 (D-2) \left(11 D^2-45 D+44\right) q^8 r_0^{24}-(D-6) (3 D-5)\cr
&&\times (5 D-11)(3 D^2-15 D+16)  r_0^{8 D}+6 (D-3) (5 D-11) \left(7 D^3-34 D^2+44 D-12\right)  \cr
&&\ \ \times q^4 r_0^{4 (D+3)}-6 (D-3)^2 (D-2) \left(45 D^2-157 D+132\right)  q^6 r_0^{2 (D+9)}\cr
&&\ \ +6 (D-2) (5 D-11) (D^3-20 D^2+79 D-80)  q^2 r_0^{6 D+6}
\Big]\,,
\nn\\
&&T_{2,2\gamma_{1}+\gamma_{2}}=24 (D-3)^2 (D-2)^2 (3 D-5) q^4 r_0^{12}\Big[(D-2) (5 D-11)  r_0^{4 D}\cr
&&\ \ +(D-2) (11 D-29) q^4 r_0^{12}-4 \left(5 D^2-24 D+29\right)  q^2 r_0^{2 D+6}
\Big]\,,
\nn\\
&&T_{2,\gamma_3}=16 (D-3)^2 (D-2) q^2 r_0^6\Big[3 (D-2) \left(23 D^3-158 D^2+357 D-262\right) q^6 r_0^{18}\cr
&&\ \ -(D-2) (D-1) D (5 D-11)  r_0^{6 D}-(135 D^4-1190 D^3+3937 D^2-5770 D\cr
&&\ \ +3144)  q^4 r_0^{2 (D+6)}+(D-2) (5 D-11) \left(11 D^2-41 D+42\right)  q^2 r_0^{4 D+6}
\Big]\,,
\nn\\
&&T_{2,\gamma_4}=16 (D-3)^2 (D-2) q^2 r_0^6\Big[3 (D-2) \left(23 D^3-160 D^2+364 D-269\right) q^6 r_0^{18}\cr
&&\ \ -\left(135 D^4-1205 D^3+4012 D^2-5900 D+3228\right)  q^4 r_0^{2 (D+6)}\cr
&&\ \ -(D-2)^2 D (5 D-11)  r_0^{6 D}+(D-2) (5 D-11) \left(11 D^2-43 D+39\right)  q^2 r_0^{4 D+6}
\Big]\,,
\nn\\
&&T_{2,\alpha_3^2}=\Big[(D-3)^2 (D-2)(1182 D^7-16091 D^6+91188 D^5-274591 D^4+460332 D^3\cr
&&\ \ -399528 D^2+126240 D+16896) q^{12} r_0^{36}+(D-4)^2 (D-2)^2 (3 D-7)^2 (3 D-5) (5 D\cr
&&\ \ -11) \left(4 D^2-23 D+26\right)  r_0^{12 D}-4 (D-3) (D-2) (3 D-7) (45 D^7-2016 D^6+24541 D^5\cr
&&\ \ -140262 D^4+438472 D^3-773928 D^2+725856 D-281600) q^6 r_0^{6 (D+3)}+(D-3)\cr
&&\ \ \times (6552 D^9-136683 D^8+1276589 D^7-6990855 D^6+24671887 D^5-58023058 D^4\cr
&&\ \ +90659368 D^3-90471808 D^2+52162048 D-13195776)  q^8 r_0^{4 (D+6)}\cr
&&\ \ -2 (D-3)^2 (D-2) (2727 D^7-39972 D^6 +250987 D^5-872306 D^4+1802632 D^3\cr
&&\ \ -2197296 D^2+1445568 D-388608)  q^{10} r_0^{2 D+30}-(D-2)^3 (3 D-7) (5 D-11) (174 D^5\cr
&&\ \ -1285 D^4+842 D^3+13773 D^2-37680 +28208)  q^4 r_0^{8 D+12}-2 (D-4) (D-2)^3\cr
&&\ \ \times (3 D-7)^2 (5 D-11) (7 D^3-131 D^2+486 D-440)  q^2 r_0^{10 D+6}\Big],\cr
&&T_{2,\alpha_3\alpha_4}=4 (D-3) (D-2) q^2 r_0^6\Big[(D-3) (D-2)^2 (3147 D^5-36169 D^4+165597 D^3\cr
&&\ \ -377075 D^2+426552 D-191556) q^{10} r_0^{30}+(D-4) (D-2)^2 (3 D-7)^2 (5 D-11) (5 D^2\cr
&&\ \ -19 D+20)  r_0^{10 D}-2 (3 D-7) (1995 D^7-32703 D^6+229484 D^5-893391 D^4\cr
&&\ \ +2083263 D^3 -2908596 D^2+2250152 D-743616)  q^4 r_0^{6 (D+2)}-(D-3) (13599 D^7\cr
&&\ \ -212199 D^6 +1416831 D^5 -5244997 D^4+11621834 D^3-15407900 D^2+11312752 D\cr
&&\ \ -3547136)  q^8 r_0^{2 D+24}+2 (D-2) (10161 D^7-169266 D^6+1205689 D^5-4758675 D^4\cr
&&\ \ +11234376 D^3-15855539 D^2 +12378446 D-4120520)  q^6 r_0^{4 D+18}+(D-2)^2\cr
&&\ \ \times (3 D-7)(5 D-11) (93 D^4-910 D^3+3363 D^2-5534 D+3420)  q^2 r_0^{8 D+6}\Big]\,,
\nn\\
&&T_{2,\alpha_4^2}=4 (D-3)^2 (D-2)^2 q^2 r_0^6\Big[(D-2)(5997 D^5-68714 D^4+313454 D^3\cr
&&\ \ -711156 D^2+801829 D-359074) q^{10} r_0^{30}-8 (D-2)^2 D (3 D-7)^2 (5 D-11)  r_0^{10 D}\cr
&&\ \ -4 (3 D-7) (1950 D^5-21187 D^4+91795 D^3-198145 D^2+212955 D-91088)  q^4 r_0^{6 (D+2)}\cr
&&\ \ -4 (6156 D^6-82959 D^5+464456 D^4-1382474 D^3+2306808 D^2-2045255 D\cr
&&\ \ +752468)  q^8 r_0^{2 D+24}+2 (18261 D^6-244728 D^5+1363126 D^4-4038156 D^3+6708433 D^2\cr
&&\ \ -5923436 D+2170932) k^2 q^6 r_0^{4 D+18}+(D-2) (3 D-7)^2 (5 D-11) (125 D^2-461 D\cr
&&\ \ +398) q^2 r_0^{8 D+6}
\Big]\,,
\nn\\
&&T_{2,\alpha _3 (2 \alpha _5+\alpha _6)}=-2 (3 D-7) (195 D^5-2423 D^4+12016 D^3-29790 D^2+36976 D\cr
&&\ \ -18400) q^2 r_0^6+(D-3) (D-2)(352 D^4-3357 D^3+11999 D^2-19068 D\cr
&&\ \ +11376) q^8 r_0^{24-6 D} -2 (D-3) (747 D^5-8697 D^4+40510 D^3-94410 D^2+110144 D\cr
&&\ \ -51488) q^6 r_0^{18-4 D}+2 (1071 D^6-15719 D^5+95953 D^4-311950 D^3+569873 D^2\cr
&&\ \ -554802 D +224936) q^4 r_0^{12-2 D}+(D-4) (D-2)^2 (3 D-7) (5 D-11) (6 D-11) r_0^{2 D}\,,
\nn\\
&&T_{2,\alpha _4 (2 \alpha _5+\alpha _6)}=
(D-2) \left(587 D^3-4264 D^2+10321 D-8324\right) q^8 r_0^{24}+6 (D-2) (3 D-7)^2\cr
&&\ \ \times (5 D-11) r_0^{8 D} +\left(3267 D^4-30572 D^3+107301 D^2-167416 D+97980\right) q^4 r_0^{4 (D+3)}\cr
&&\ \ -\left(2349 D^4-21960 D^3+76955 D^2-119816 D+69936\right) q^6 r_0^{2 (D+9)}\cr
&&\ \ -(3 D-7) \left(605 D^3-4219 D^2+9828 D-7648\right) q^2 r_0^{6 D+6}\,,
\nn\\
&&T_{2, (2 \alpha _5+\alpha _6)^2}=-\left(816 D^3-5679 D^2+13179 D-10198\right) q^2 r_0^{6-2 D}+2 (321 D^3-2207 D^2\cr
&&\ \ +5060 D-3868) q^4 r_0^{12-4 D}-(D-2) \left(166 D^2-783 D+923\right) q^6 r_0^{18-6 D}\cr
&&\ \ +8 (3 D-7)^2 (5 D-12)\ .
\eea
The entropy computed using Wald formalism is given by
\bea
S'&=&\frac{\omega_{D-2}}{4}r_0^{D-2}\left(1+\Delta {S}_1+\Delta{S}_2\right)\,,
\nn\\
\Delta {S}_1&=&\frac{r_0^{-2 D-2}}{(3 D-7) \left( r_0^{2 D}-q^2 r_0^6\right)}\Big[\Big(
(D-3) \left(7 D^2-36 D+48\right) q^4 r_0^{12}+(D-2)^2 (3 D-7)  r_0^{4 D}\cr
&&-2 (D-3) (D-2) (3 D-7)  q^2 r_0^{2 D+6}
\Big)\alpha_3+2 (D-3) (D-2) q^2 r_0^6 \big((5 D-13) q^2 r_0^6\cr
&&-(3 D-7)  r_0^{2 D}\big)\alpha_4+8  (D-3) (D-2)^2 q^4 r_0^{12}(2 \alpha _5+\alpha _6)
\Big]\,,
\nn\\
\Delta {S}_2&=&-\frac{S_{2,\beta_7}\beta_{7}+S_{2,\beta_8}\beta_{8}+S_{2,2 \gamma_{1}+\gamma_{2}}\left(2 \gamma_{1}+\gamma_{2}\right)+S_{2,\gamma_3}\gamma_{3}+S_{2,\gamma_4}\gamma_{4}
}{8 (3 D-5) (5 D-11) r_0^{4 (D+1)}\left( r_0^{2 D}-q^2 r_0^6\right)}\cr
&&-\frac{S_{2,\alpha_3^2}\alpha_3^2+S_{2,\alpha_3\alpha_4}\alpha_3\alpha_4+S_{2,\alpha_4^2}\alpha_4^2
}{2 (D-2) (3 D-7)^2 (3 D-5) (5 D-11)r_0^{4 (D+1)} \left( r_0^{2 D}-q^2 r_0^6\right){}^3}\\
&&+\frac{8  \left(2 \alpha _5+\alpha _6\right) (D-3) (D-2) q^4 }{(3 D-7)^2 (5 D-11)r_0^{4 D-8} \left(r_0^{2 D}-q^2 r_0^6\right){}^3}\Big[S_{2,\alpha _3 (2 \alpha _5+\alpha _6)}\alpha _3\cr
&&-2 (D-3) (D-2)S_{2,\alpha _4 (2 \alpha _5+\alpha _6)} \alpha _4+4 (D-3) (D-2)^2 q^2 r_0^6S_{2, (2 \alpha _5+\alpha _6)^2}\left(2 \alpha _5+\alpha _6\right)\Big],\nn
\eea
where
\bea
&&S_{2,\beta_7}=4\Big[(D-3)^2 (129 D^4-1239 D^3+4410 D^2-6886 D+3964) q^6 r_0^{18}-(D-2)
 \cr
&&\ \ \times (3 D-5) (5 D-11) \left(D^3-9 D^2+26 D-22\right) r_0^{6 D}-3 (D-3)^2 (D-2) (5 D-11) \cr
&&\ \ \times \left(11 D^2-54 D+62\right)  q^4 r_0^{2 (D+6)} +3 (D-3) (D-2) (5 D-11)(5 D^3-37 D^2\cr
&&\ \ +86 D-62)  q^2 r_0^{4 D+6}\Big]\,,
\nn\\
&&S_{2,\beta_8}=(D-2)\Big[3 (D-3)^2 \left(11 D^2-45 D+44\right) q^6 r_0^{18}-(3 D-5) (5 D-11) (3 D^2\cr
&&\ \ -15 D+16)  r_0^{6 D}-3 (D-3)^2 (5 D-11) (7 D-12)  q^4 r_0^{2 (D+6)}+3 (D-3) (5 D-11) \cr
&&\ \ \times \left(7 D^2-31 D+32\right)  q^2 r_0^{4 D+6}\Big]\,,
\nn\\
&&S_{2,2 \gamma_{1}+\gamma_{2}}=8 (D-3)^2 (D-2)^2 (3 D-5) q^4 r_0^{12}  \left((11 D-29) q^2 r_0^6-(5 D-11)  r_0^{2 D}\right)\,,
\nn\\
&&S_{2,\gamma_3}=16 (D-3)^2 (D-2) q^2 r_0^6\Big[\left(23 D^3-158 D^2+357 D-262\right) q^4 r_0^{12}\cr
&&\ \ +(D-2) (D-1) (5 D-11)  r_0^{4 D}-4 (D-2)^2 (5 D-11)  q^2 r_0^{2 D+6}
\Big]\,,
\nn\\
&&S_{2,\gamma_4}=16 (D-3)^2 (D-2) q^2 r_0^6\Big[\left(23 D^3-160 D^2+364 D-269\right) q^4 r_0^{12}\cr
&&\ \ +(D-2)^2 (5 D-11)  r_0^{4 D}-(5 D-11) \left(4 D^2-17 D+17\right)  q^2 r_0^{2 D+6}
\Big]\,,
\nn\\
&&S_{2,\alpha_3^2}=\Big[-(D-3)^2 (96 D^7-2551 D^6+24438 D^5-119535 D^4+333660 D^3\cr
&&\ \ -540296 D^2+474016 D-174592) q^{10} r_0^{30}+(D-4)^2 (D-2)^2 (2 D-3) (3 D-7)^2\cr
&&\ \ \times (3 D-5) (5 D-11)  r_0^{10 D} +2 (D-3) (D-2) (3 D-7) (5 D-11) (102 D^5-1197 D^4\cr
&&\ \ +5379 D^3-11450 D^2+11288 D -3904)  q^4 r_0^{6 (D+2)}-(D-3)^2 (18 D^7+1125 D^6\cr
&&\ \ -15208 D^5+78313 D^4-207940 D^3 +302264 D^2-226912 D+67584)  q^8 r_0^{2 D+24}\cr
&&\ \ -2 (D-3) (D-2) (3 D-7) (156 D^6-2021 D^5 +9596 D^4-18497 D^3+3450 D^2\cr
&&\ \ +33144 D-30976)  q^6 r_0^{4 D+18}-(D-4) (D-2)^2 (3 D-7)^2(5 D-11) (40 D^3-271 D^2\cr
&&\ \ +573 D-372)  q^2 r_0^{8 D+6}\Big]\,,
\nn\\
&&S_{2,\alpha_3\alpha_4}=4 (D-3)^2 (D-2) q^2 r_0^6\Big[(D-2)^2 (699 D^4-6595 D^3+23167 D^2-35805 D\cr
&&\ \ +20486) q^8 r_0^{24} +(D-4) (D-2)^2 (3 D-7)^2 (5 D-11)  r_0^{8 D}+2 (3 D-7) (504 D^5\cr
&&\ \ -5770 D^4+26387 D^3-60217 D^2+68526 D-31084)  q^4 r_0^{4 (D+3)}-2 (1269 D^6\cr
&&\ \ -17205 D^5+96945 D^4-290407 D^3 +487530 D^2-434736 D+160836)  q^6 r_0^{2 (D+9)}\cr
&&-2 (D-3) (D-2) (3 D-7) (5 D-11) (33 D^2 -144 D+158)  q^2 r_0^{6 D+6}
\Big]\,,
\nn\\
&&S_{2,\alpha_4^2}=4 (D-3)^2 (D-2)^2 q^2 r_0^6\Big[
(1749 D^5-20138 D^4+92318 D^3\cr
&&\ \ -210484 D^2+238477 D-107298) q^8 r_0^{24}+8 (D-2)^2 (3 D-7)^2 (5 D-11)  r_0^{8 D}\cr
&&\ \ +(3 D-7) (3 D-5) (631 D^3-4582 D^2+11085 D-8934)  q^4 r_0^{4 (D+3)}\cr
&&\ \ -\left(5283 D^5-60138 D^4+272498 D^3-614004 D^2+687467 D-305698\right)  q^6 r_0^{2 (D+9)}\cr
&&\ \ -(3 D-7)^2 (5 D-11) \left(53 D^2-221 D+222\right)  q^2 r_0^{6 D+6}
\Big]\,,
\nn\\
&&S_{2,\alpha _3 (2 \alpha _5+\alpha _6)}=(D-4) (D-2) (3 D-7) (4 D-9) (5 D-11) r_0^{6 D}-(D-3) (94 D^4\cr
&&\ \ -901 D^3+3233 D^2-5156 D+3088) q^6 r_0^{18}+(D-3) (288 D^4-2737 D^3+9719 D^2\cr
&&\ \ -15288 D +8992) q^4 r_0^{2 (D+6)}-(3 D-7) (98 D^4-1009 D^3+3867 D^2-6554 D\cr
&&\ \ +4152) q^2 r_0^{4 D+6}\,,
\nn\\
&&S_{2,\alpha _4 (2 \alpha _5+\alpha _6)}=
(3 D-7) (133 D^2-637 D+762) q^2 r_0^{4 D+6}+(179 D^3-1308 D^2\cr
&&\ \ +3185 D-2584) q^6 r_0^{18} -2 (3 D-7)^2 (5 D-11) r_0^{6 D}-4 (117 D^3-842 D^2\cr
&&\ \ +2018 D-1611) q^4 r_0^{2 (D+6)}\,,
\nn\\
&&S_{2, (2 \alpha _5+\alpha _6)^2}=\left(114 D^2-531 D+619\right) q^2 r_0^{2 D+6}+\left(-52 D^2+247 D-293\right) q^4 r_0^{12}\cr
&&\ \ -8 (3 D-7)^2 r_0^{4 D}.
\eea
Finally we present the electric potential which is given by the value of temperal component of gauge field at the horizon
\bea
\Psi_e'&=&\sqrt{\frac{2(D-2)}{D-3}}\frac{ q}{r_0^{D-3}}+\Delta\Psi'_{e,1}+\Delta\Psi'_{e,2}\,,
\nn\\
%%%
\Delta\Psi'_{e,1}&=&\frac{\sqrt{2} \sqrt{D-3} q r_0^{1-3 D}}{\sqrt{D-2} (3 D-7)( r_0^{2 D}-q^2 r_0^6)}\Big[\Big((D-3) (7 D^2-36 D+48) q^4 r_0^{12}\cr
&&+(D-4) (D-2) (3 D-7)  r_0^{4 D}-2 (D-3) (7 D^2-36 D+48)  q^2 r_0^{2 D+6}
\Big)\alpha_3\cr
&&+2 (D-3) (D-2)\Big((3 D-7) r_0^{4 D}-2 (5 D-13)  q^2 r_0^{2 D+6}+(5 D-13) q^4 r_0^{12}
\Big)\alpha_4\cr
&&+8  (D-3) (D-2)^2 q^2 r_0^6 (q^2 r_0^6-2 r_0^{2 D})(2 \alpha _5+\alpha _6)
\Big]\,,
\nn\\
\Delta\Psi'_{e,2}&=&-\frac{\sqrt{D-3} q r_0^{-5 D-1}\big(\Psi_{2,\beta_7}\beta_{7}+\Psi_{2,\beta_8}\beta_{8}+\Psi_{2,2\gamma_1+\gamma_2}\left(2 \gamma_{1}+\gamma_{2}\right)+\Psi_{2,\gamma_3}\gamma_{3}+\Psi_{2,\gamma_4}\gamma_{4}
\big)}{2 \sqrt{2} \sqrt{D-2} (3 D-5) (5 D-11) \left( r_0^{2 D}-q^2 r_0^6\right)}\cr
&&-\frac{\sqrt{D-3} q r_0^{-5 D-1}\big(\Psi_{2,\alpha_3^2}\alpha_3^2+\Psi_{2,\alpha_3\alpha_4}\alpha_3\alpha_4+\Psi_{2,\alpha_4^2}\alpha_4^2
\big)}{\sqrt{2} (D-2)^{3/2} (3 D-7)^2 (3 D-5) (5 D-11) \left( r_0^{2 D}-q^2 r_0^6\right){}^3}\cr
&&+\frac{16 \sqrt{2}  \left(2 \alpha _5+\alpha _6\right) (D-3)^{3/2} \sqrt{D-2} q^3 r_0^{5-5 D}}{(3 D-7)^2 (5 D-11) \left(r_0^{2 D}-q^2 r_0^6\right){}^3}\big[\Psi_{2,\alpha _3 (2 \alpha _5+\alpha _6)}\alpha _3\nn\\
&&-2 (D-3) (D-2)\Psi_{2,\alpha _4 (2 \alpha _5+\alpha _6)}\alpha _4 +4 (D-3) (D-2)^2 q^2 r_0^6\Psi_{2,(2 \alpha _5+\alpha _6)^2}\left(2 \alpha _5+\alpha _6\right)\big]\ .
\nn
\eea
where
\bea
&&\Psi_{2,\beta_7}=4\Big[(D-3)^2 (129 D^4-1239 D^3+4410 D^2-6886 D+3964) q^6 r_0^{18} -2 (D-2)(5 D\cr
&&\ \ -11) \left(3 D^4-31 D^3+117 D^2-191 D+112\right)  r_0^{6 D} -6 (D-3)^2 (46 D^4-435 D^3+1524 D^2\cr
&&\ \ -2344 D+1332)  q^4 r_0^{2 (D+6)}+3 (D-3)^2 (D-2) (5 D-11) \left(11 D^2-54 D+62\right)  q^2 r_0^{4 D+6}
\Big]\,,
\nn\\
&&\Psi_{2,\beta_8}=(D-2)\Big[3 (D-3)^2 \left(11 D^2-45 D+44\right) q^6 r_0^{18}-2 (5 D-11) (3 D^3-24 D^2+63 D\cr
&&\ \ -52) r_0^{6 D}-6 (D-3)^2 \left(17 D^2-68 D+66\right)  q^4 r_0^{2 (D+6)}\cr
&&\ \ +3 (D-3)^2 (5 D-11) (7 D-12)  q^2 r_0^{4 D+6}
\Big]\,,
\nn\\
&&\Psi_{2,2\gamma_1+\gamma_2}=8 (D-3)^2 (D-2)^2 (3 D-5) q^2 r_0^6  \big[(5 D-11)  r_0^{4 D}-(19 D-49)  q^2 r_0^{2 D+6}\cr
&&\ \ +(11 D-29) q^4 r_0^{12}\big]\,,
\nn\\
&&\Psi_{2,\gamma_3}=8 (D-3)^2 (D-2)\Big[2 \left(23 D^3-158 D^2+357 D-262\right) q^6 r_0^{18}-(D-2) (D-1) (5 D\cr
&&\ \ -11)  r_0^{6 D}-\left(89 D^3-598 D^2+1327 D-962\right) q^4 r_0^{2 (D+6)}+8 (D-2)^2 (5 D-11)  q^2 r_0^{4 D+6}
\Big]\,,
\nn\\
&&\Psi_{2,\gamma_4}=8 (D-3)^2 (D-2)\Big[2 \left(23 D^3-160 D^2+364 D-269\right) q^6 r_0^{18}\cr
&&\ \ -(D-2)^2 (5 D-11)  r_0^{6 D}-\left(89 D^3-609 D^2+1364 D-994\right)  q^4 r_0^{2 (D+6)}\cr
&&\ \ +2 (5 D-11) \left(4 D^2-17 D+17\right)  q^2 r_0^{4 D+6}
\Big]\,,
\nn\\
&&\Psi_{2,\alpha_3^2}=2\Big[(D-3)^2 (3 D-7) (213 D^6-2610 D^5+13181 D^4-34932 D^3+50824 D^2\cr &&\ \ -38048 D +11264) q^{10} r_0^{30}+2 (D-4) (D-2)^2 (3 D-7)^2 (5 D-11)(7 D^3-47 D^2\cr
&&\ \ +102 D-68)  r_0^{10 D}\cr
&&\ \ +4 (D-3) (D-2) (3 D-7)(12 D^6-51 D^5-1101 D^4+10489 D^3-36253 D^2\cr
&&\ \ +56332 D -33088)  q^4 r_0^{6 (D+2)}-2 (D-3)^2 (1407 D^7-21482 D^6+141521 D^5\cr
&&\ \ -521512 D^4+1160550 D^3-1558600 D^2+1168800 D-377344)  q^8 r_0^{2 D+24}\cr
&&\ \ +2 (D-3)^2 (3 D-7) (531 D^6-7106 D^5 +40763 D^4-128056 D^3+231192 D^2\cr
&&\ \ -226016 D+92928)  q^6 r_0^{4 D+18}-(D-2) (3 D-7)^2 (5 D-11) (45 D^5-562 D^4\cr
&&\ \ +2707 D^3-6190 D^2+6512 D-2336)  q^2 r_0^{8 D+6}
\Big]\,,
\nn\\
&&\Psi_{2,\alpha_3\alpha_4}=8 (D-3) (D-2)\Big[4 (D-3) (D-2) (3 D-7) (102 D^4-936 D^3+3201 D^2\cr
&&\ \ -4820 D+2689) q^{10} r_0^{30}+(D-4) (D-2)^2 (D-1) (3 D-7)^2 (5 D-11)  r_0^{10 D}\cr
&&\ \ -2 (3 D-7)(771 D^6-11126 D^5+66757 D^4-213092 D^3+381452 D^2-362830 D\cr
&&\ \ +143156)  q^4 r_0^{6 (D+2)}-(D-3)(5259 D^6-71711 D^5+406849 D^4-1228537 D^3\cr
&&\ \ +2081204 D^2-1874240 D+700536)  q^8 r_0^{2 D+24}+4 (D-3) (3 D-7) (648 D^5\cr
&&\ \ -7363 D^4 +33437 D^3-75751 D^2+85497 D-38408)  q^6 r_0^{4 D+18}\cr
&&\ \ +4 (D-2) (3 D-7)^2 (5 D-11) (4 D^3 -32 D^2+86 D-77)  q^2 r_0^{8 D+6}
\Big]\,,
\nn\\
&&\Psi_{2,\alpha_4^2}=16 (D-3)^2 (D-2)^2\Big[2 (3 D-7)(177 D^4-1611 D^3+5455 D^2-8133 D+4496)
\cr
&&\ \ \times q^{10} r_0^{30}-(3 D-7) \left(1386 D^4-12379 D^3+41153 D^2-60269 D+32749\right) q^4 r_0^{6 (D+2)}\cr
&&\ \ +\left(-4332 D^5+49517 D^4-225346 D^3+510048 D^2-573714 D+256307\right)  q^8 r_0^{2 D+24}\cr
&&\ \ +8 (3 D-7)^2 \left(89 D^3-597 D^2+1314 D-944\right)  q^6 r_0^{4 D+18}+4 (2 D-5) (3 D-7)^2 \cr
&&\ \ \times (3 D-5) (5 D-11) q^2 r_0^{8 D+6}-2 (D-2)^2 (3 D-7)^2 (5 D-11)  r_0^{10 D}
\Big]\,,
\nn\\
&&\Psi_{2,\alpha _3 (2 \alpha _5+\alpha _6)}=
-4 (D-3) (3 D-7) (63 D^3-458 D^2+1111 D-900) q^4 r_0^{4 (D+3)}+(D-3)\cr
&&\ \ \times \left(539 D^4-5168 D^3+18601 D^2-29800 D+17936\right) q^6 r_0^{2 (D+9)}+(3 D-7) (137 D^4\cr
&&\ \ -1414 D^3+5445 D^2-9284 D+5920) q^2 r_0^{6 D+6}-(D-3) (3 D-7) (43 D^3-309 D^2\cr
&&\ \ +740 D-592) q^8 r_0^{24}-(D-4) (D-2) (3 D-7)^2 (5 D-11) r_0^{8 D}\,,
\nn\\
&&\Psi_{2,\alpha _4 (2 \alpha _5+\alpha _6)}=
2 (3 D-7) \left(181 D^2-882 D+1073\right) q^4 r_0^{4 (D+3)}\cr
&&\ \ -\left(799 D^3-5790 D^2+13979 D-11244\right) q^6 r_0^{2 (D+9)}\cr
&&\ \ -(3 D-7) \left(197 D^2-945 D+1132\right) q^2 r_0^{6 D+6}\cr
&&\ \ +2 (2 D-5) (3 D-7) (17 D-41) q^8 r_0^{24}+2 (3 D-7)^2 (5 D-11) r_0^{8 D}\,,
\nn\\
&&\Psi_{2,(2 \alpha _5+\alpha _6)^2}=
2 \left(106 D^2-499 D+587\right) q^4 r_0^{2 (D+6)}-2 (3 D-7) (43 D-101) q^2 r_0^{4 D+6}\cr
&&\ \ +(3 D-7) (45-19 D) q^6 r_0^{18}+12 (3 D-7)^2 r_0^{6 D}\ .
\eea

\section{Redefinition of parameters for the charged solution}
\label{app:redefcharge}

In order to compare results obtained by our method \eqref{onshell1} to those obtained using ordinary method, we need to redefine the parameters $r_0,\, q$ so that in terms of the new parameters, the temperature and potential retain the same functional form at a required order of small parameters. At the first and second order in $\alpha_i$, this is achieved by doing the shift
\bea
&&\tilde{r}_0\rightarrow r_0- \delta_1r_0-\Big[\delta_2r_0+\frac{(\delta_1r_0)^2}{r_0}+\frac{2 \left(D^2-7 D+12\right) q^2 r_0^{3-4 D}}{(D-2)^2 (3 D-7)^2 \left(r_0^{2 D}-q^2 r_0^6\right){}^3}\cr
&&~\times\Big( \big((D-3) (7 D^2-36 D+48) q^4 r_0^{12}+(D-4) (D-2) (3 D-7) r_0^{4 D}\cr
&&~-2 (D-3) (7 D^2-36 D+48) q^2 r_0^{2 D+6}\big)\alpha _3+2  (D-3) (D-2) \big((5 D-13) q^4 r_0^{12}\cr
&&~+(3 D-7) r_0^{4 D}-2 (5 D-13) q^2 r_0^{2 D+6}\big)\alpha _4\Big)^2
\Big],\cr
&&\tilde{q}\rightarrow q- \delta_1q-\Big[\delta_2q-\frac{(D-3) q r_0^{-4 (D+1)}}{(D-2)^2 (3 D-7)^2 \left(r_0^{2 D}-q^2 r_0^6\right){}^3}\Big(
(2 (D-3) (57 D^8-1114 D^7\cr
&&~+9318 D^6-43252 D^5+120761 D^4-205242 D^3+204136 D^2-106624 D\cr
&&~+22272) q^4 r_0^{6 (D+2)}+(D-3)^2 \left(7 D^2-36 D+48\right)(19 D^5-225 D^4+1023 D^3-2117 D^2\cr
&&~+1708 D-80) q^8 r_0^{2 D+24}-2 (D-3) (81 D^8-1524 D^7+12056 D^6-50930 D^5\cr
&&~+117887 D^4 -122494 D^3-36680 D^2+204416 D-135936) q^6 r_0^{4 D+18}-(D-4) (D-2)\cr
&&~\times(3 D-7)(15 D^6-220 D^5+1294 D^4 -3866 D^3+6131 D^2-4890 D+1592) q^2 r_0^{8 D+6}\cr
&&~-(D-3)^3 \left(D^2-6 D+11\right)\left(7 D^2-36 D+48\right)^2 q^{10} r_0^{30}+
(D-4)^2 (D-2)^2 (D-1)\cr &&~\times(3 D-7)^2\left(D^2-6 D+7\right) r_0^{10 D})\alpha _3^2
+4(D-3) (D-2) (2 (21 D^7-289 D^6+1454 D^5\cr
&&~-2508 D^4-4075 D^3+23598 D^2-35361 D +18300) q^4 r_0^{6 (D+2)}+(D-3) (86 D^6\cr
&&~-1213 D^5+6914 D^4 -19848 D^3+28525 D^2-16012 D -752) q^8 r_0^{2 D+24}-2 (42 D^7\cr
&&~-611 D^6+3326 D^5-6835 D^4-6227 D^3+55248 D^2-95691 D +56948) q^6 r_0^{4 D+18}\cr
&&~-(D-3) (3 D-7) \left(3 D^5-21 D^4+15 D^3+177 D^2-410 D+216\right) q^2 r_0^{8 D+6}
-(D-3)^2\cr
&&~\times (5 D-13) \left(D^2-6 D+11\right) \left(7 D^2-36 D+48\right) q^{10} r_0^{30}+(D-4) (D-2) (3 D-7)^2\cr
&&~\times \left(D^2-5 D+5\right) r_0^{10 D})\alpha _3 \alpha _4
\Big)-4  (D-3)^2 (D-2)^2 (-(9 D^5+69 D^4-1335 D^3\cr
&&~+6025 D^2-11110 D+7398) q^4 r_0^{6 (D+2)}-(5 D-13)(11 D^4-94 D^3+268 D^2-236 D\cr
&&~-53) q^8 r_0^{2 D+24}+(39 D^5-217 D^4-613 D^3+6907 D^2-17126 D+13890) q^6 r_0^{4 D+18}\cr
&&~+(D-1) (3 D-7) \left(12 D^2-69 D+101\right) q^2 r_0^{8 D+6}+(D-3) (5 D-13)^2\cr
&&~\times \left(D^2-6 D+11\right) q^{10} r_0^{30}-(-7+3 D)^2 (D-3) r_0^{10 D})\alpha _4^2
\Big]\,.
\label{parameter redefinition}
\eea
The first order shifts of parameters are
\bea
\delta_1r_0&=&-\frac{(D-4) r_0^{-2 D-1}}{(D-2) (3 D-7) \left( r_0^{2 D}-q^2 r_0^6\right)}\Big[\big(
(D-3) \left(7 D^2-36 D+48\right) q^4 r_0^{12}\cr
&&+(D-2)^2 (3 D-7)  r_0^{4 D}-2 (D-3) (D-2) (3 D-7)  q^2 r_0^{2 D+6}
\big)\alpha_3\cr
&&+2 (D-3) (D-2) q^2 r_0^6  \left((5 D-13) q^2 r_0^6-(3 D-7)  r_0^{2 D}\right)\alpha_4\cr
&&+8 (D-3) (D-2)^2 q^4 r_0^{12}\left(2 \alpha _5+\alpha _6\right)
\Big]\,,
\nn\\
\delta_1q&=&-\frac{(D-3) q r_0^{-2 D-2}}{(D-2) (3 D-7) \left( r_0^{2 D}-q^2 r_0^6\right)}\Big[\big(
(D-3)^2 \left(7 D^2-36 D+48\right) q^4 r_0^{12}\cr
&&+(D-4) (D-2) (D-1) (3 D-7)  r_0^{4 D}-2 (D-3) \left(3 D^3-18 D^2+30 D-8\right)\cr
&& \ \ \times q^2 r_0^{2 D+6}\big)\alpha_3 +2 (D-3) (D-2)\big((3 D-7)  r_0^{4 D}+(D-3) (5 D-13) q^4 r_0^{12}\cr
&&\ \ -\left(3 D^2-9 D+2\right)  q^2 r_0^{2 D+6}
\big)\alpha_4-8  (D-3) (D-2)^2 q^2 r_0^6 (-D q^2 r_0^6+2 r_0^{2 D}\cr
&&\ \ +3 q^2 r_0^6)\left(2 \alpha _5+\alpha _6\right)
\Big]\,,
\eea
and the second order shifts of parameters are
\bea
\delta_2r_0&=&\frac{
r_{0,\beta_7}\beta_{7}+r_{0,\beta_8}\beta_{8}+ r_{0,2\gamma_1+\gamma_2}\left(2 \gamma_{1}+\gamma_{2}\right)+r_{0,\gamma_3}\gamma_{3}+r_{0,\gamma_4}\gamma_{4}}{8 (D-2) (3 D-5) (5 D-11)r_0^{4 D+3} \left( r_0^{2 D}-q^2 r_0^6\right)}\cr
&&+\frac{r_{0,\alpha_3^2}\alpha_3^2+r_{0,\alpha_3\alpha_4}\alpha_3\alpha_4+r_{0,\alpha_4^2}\alpha_4^2}{2 (D-2)^2 (3 D-7)^2 (3 D-5) (5 D-11)r_0^{4 D+3} \left( r_0^{2 D}-q^2 r_0^6\right){}^3}\cr
&&+\frac{8 \left(2 \alpha _5+\alpha _6\right) (D-3) q^4 r_0^9}{(3 D-7)^2 (5 D-11) r_0^{4 D} \left( r_0^{2 D}-q^2 r_0^6\right){}^3}\big(r_{0, \alpha _3(2 \alpha _5+\alpha _6)} \alpha _3+(D-3) (D-2)\cr
&&\ \ \times r_{0, \alpha _4(2 \alpha _5+\alpha _6)}\alpha _4+4 (D-3) (D-2)^2 q^2 r_0^6r_{0, (2 \alpha _5+\alpha _6)^2}\left(2 \alpha _5+\alpha _6\right)\big)\,,
\nn\\
\delta_2q&=&-\frac{(D-3) q\big(q_{\beta_7}\beta_{7}+q_{\beta_8}\beta_{8}+q_{2\gamma_1+\gamma_2}(2 \gamma_{1}+\gamma_{2})+q_{\gamma_3}\gamma_{3}+q_{\gamma_4}\gamma_{4}
\big) }{8 (D-2) (3 D-5) (5 D-11) r_0^{4 D+4}\left(q^2 r_0^6-r_0^{2 D}\right)}\cr
&&+\frac{(D-3) q \big(q_{\alpha_3^2}\alpha_3^2+q_{\alpha_3\alpha_4}\alpha_3\alpha_4+q_{\alpha_4^2}\alpha_4^2
\big)}{2 (D-2)^2 (3 D-7)^2 (3 D-5) (5 D-11)r_0^{4 D+4} \left(q^2 r_0^6- r_0^{2 D}\right){}^2}\cr
&&+\frac{8 \left(2 \alpha _5+\alpha _6\right) (D-3)^2 q^3 r_0^{2-4 D}}{(3 D-7)^2 (5 D-11) \left(q^2 r_0^6-r_0^{2 D}\right){}^2}\big(q_{\alpha _3(2 \alpha _5+\alpha _6) } \alpha _3+2 (D-3) (D-2)q_{\alpha _4(2 \alpha _5+\alpha _6) }\alpha _4\cr
&&+4 (D-3) (D-2)^2 q^2 r_0^6q_{(2 \alpha _5+\alpha _6)^2 }\left(2 \alpha _5+\alpha _6\right)\big).\label{rq transformation}
\eea
where
\bea
&&r_{0,\beta_7}=4 (D-6)\Big[3 (D-3) (D-2) (5 D-11) (5 D^3-37 D^2+86 D-62)  q^2 r_0^{4 D+6}\cr
&&~ -(D-2)(3 D-5) (5 D-11) \left(D^3-9 D^2+26 D-22\right)  r_0^{6 D}-3 (D-3)^2 (D-2)\cr
&&~ \times (5 D-11)\left(11 D^2-54 D+62\right)  q^4 r_0^{2 D+12}+(D-3)^2 (129 D^4-1239 D^3+4410 D^2\cr
&&~ -6886 D+3964) q^6 r_0^{18}
\Big]\,,
\nn\\
&&r_{0,\beta_8}=(D-6) (D-2)\Big[3 (D-3)^2 \left(11 D^2-45 D+44\right) q^6 r_0^{18}-(3 D-5) (5 D-11)\cr
&&~ \times \left(3 D^2-15 D+16\right)  r_0^{6 D}-3 (D-3)^2 (5 D-11) (7 D-12)  q^4 r_0^{2 (D+6)}\cr
&&~+3 (D-3) (5 D-11) \left(7 D^2-31 D+32\right)  q^2 r_0^{4 D+6}
\Big]\,,
\nn\\
&&r_{0,2\gamma_1+\gamma_2}=8 (D-6) (D-3)^2 (D-2)^2 (3 D-5) q^4 r_0^{12} \left[(11 D-29) q^2 r_0^6-(5 D-11)  r_0^{2 D}\right],\cr
&&~r_{0,\gamma_3}=16 (D-6) (D-3)^2 (D-2) q^2 r_0^6\Big[\left(23 D^3-158 D^2+357 D-262\right) q^4 r_0^{12}\cr
&&~+(D-2) (D-1) (5 D-11) r_0^{4 D}-4 (D-2)^2 (5 D-11)  q^2 r_0^{2 D+6}
\Big]\,,
\nn\\
&&r_{0,\gamma_4}=16 (D-6) (D-3)^2 (D-2) q^2 r_0^6\Big[(23 D^3-160 D^2+364 D-269) q^4 r_0^{12}\cr
&&~+(D-2)^2 (5 D-11)  r_0^{4 D}-(5 D-11) (4 D^2-17 D+17)  q^2 r_0^{2 D+6}
\Big]\,,
\nn\\
&&r_{0,\alpha_3^2}=-\Big[(D-3)^2 (96 D^8-3127 D^7+38274 D^6-245359 D^5+927976 D^4\cr
&&~-2154848 D^3+3029296 D^2-2371264 D+794112) q^{10} r_0^{30} -2 (D-3) (D-2)\cr
&&~ \times(3 D-7) (5 D-11) (72 D^6-1495 D^5+11833 D^4 -46724 D^3+98076 D^2\cr
&&~ -104016 D+43264)  q^4 r_0^{6 (D+2)}-(D-3)^2(1872 D^8-44925 D^7+458528 D^6\cr
&&~-2612065 D^5+9118706 D^4-20036688 D^3 +27121200 D^2-20702784 D\cr
&&~ +6825984)  q^8 r_0^{2 D+24}+2 (D-3) (1608 D^9-42727 D^8+490165 D^7-3204057 D^6\cr
&&~ +13202051 D^5-35657856 D^4 +63255432 D^3-71163800 D^2+46108320 D\cr
&&~ -13111296)  q^6 r_0^{4 D+18}+(D-4) (D-2)^2 (3 D-7)^2 (5 D-11)(16 D^4-279 D^3\cr
&&~ +1417 D^2-2728 D+1712)  q^2 r_0^{8 D+6}+(D-4)^2 (D-2)^2 (3 D-7)^2 (3 D-5)\cr
&&~ \times (5 D-11) (7 D-10)  r_0^{10 D}\Big]\,,
\nn\\
&&r_{0,\alpha_3\alpha_4}=4 (D-3) (D-2) q^2 r_0^6\Big[(D-3) (699 D^7-13585 D^6+109739 D^5\cr
&&~-481101 D^4+1242018 D^3-1894066 D^2+1583192 D-560304) q^8 r_0^{24}\cr
&&~+(D-4) (D-2) (3 D-7)^2 (5 D-11) (D^3+D^2-26 D+34)  r_0^{8 D}\cr
&&~-2 (D-3) (459 D^7-8487 D^6+62795 D^5-242473 D^4+525274 D^3\cr
&&~-627444 D^2+366484 D-70936)  q^6 r_0^{2 D+18}+2 (132 D^8-2140 D^7+9625 D^6\cr
&&~+19545 D^5-334587 D^4+1320245 D^3-2562010 D^2+2522870 D\cr
&&~ -1007456)  q^4 r_0^{4 D+12}-2 (3 D-7) (5 D-11) (3 D^6-4 D^5-363 D^4+2898 D^3\cr
&&~ -9142 D^2+13084 D-7016) q^2 r_0^{6 D+6}\Big],\cr
&&r_{0,\alpha_4^2}=4 (D-3)^2 (D-2)^2 q^2 r_0^6\Big[(1749 D^6-30632 D^5+213896 D^4-771192 D^3\cr
&&~ +1524281 D^2-1572064 D+662378) q^8 r_0^{24}+2 (3 D-7)^2 (5 D-11) (D^3-11 D^2\cr
&&~ +27 D-21)  r_0^{8 D}+(3 D-5) (983 D^5-15459 D^4+90005 D^3-249317 D^2+333268 D\cr
&&~ -173400)  q^4 r_0^{4 D+12}-(3 D-7) (5 D-11) (57 D^4-732 D^3+2935 D^2-4704 D\cr
&&~ +2644)  q^2 r_0^{6 D+6}-(3933 D^6-68556 D^5+472114 D^4-1670016 D^3+3228429 D^2\cr
&&-3250996 D+1336548)  q^6 r_0^{2 D+18}\Big],
\nn\\
&&r_{0, \alpha _3(2 \alpha _5+\alpha _6)}=-(D-3) (138 D^5-2115 D^4+11857 D^3-30994 D^2+37968 D\cr
&&~ -17344) q^4 r_0^{2 (D+6)}+(74 D^6-1333 D^5+9040 D^4-29655 D^3+48482 D^2-34360 D\cr
&&~ +5248) q^2 r_0^{4 D+6}+(D-3) (94 D^5-1465 D^4+8709 D^3-25068 D^2+35296 D\cr
&&~ -19584) q^6 r_0^{18}-(D-4) (D-2) (3 D-7) (5 D-11) \left(2 D^2-13 D+10\right) r_0^{6 D}\,,
\nn\\
&&~r_{0, \alpha _4(2 \alpha _5+\alpha _6)}=
2 (199 D^4-2676 D^3+12003 D^2-22388 D+14998) q^2 r_0^{4 D+6}\cr
&&~+2 \left(179 D^4-2382 D^3+11083 D^2-21934 D+15790\right) q^6 r_0^{18}\cr
&&~-8 (87 D^4-1158 D^3+5296 D^2-10215 D+7136) q^4 r_0^{2 (D+6)}\cr
&&~-4 (3 D-7) (5 D-11) \left(D^2-9 D+12\right) r_0^{6 D}\,,
\nn\\
&&r_{0, (2 \alpha _5+\alpha _6)^2}=
(52 D^3-559 D^2+1785 D-1780) q^4 r_0^{12}+8 (4 D^3-45 D^2\cr
&&~ +138 D-129) r_0^{4 D}-\left(84 D^3-909 D^2+2837 D-2746\right) q^2 r_0^{2 D+6},\cr
&&~q_{\beta_7}=4\Big[(D-4) (D-3)^2 (129 D^4-1239 D^3+4410 D^2-6886 D+3964) q^6 r_0^{18}\cr
&&~-(D-2) (D-1) (5 D-11) \left(3 D^4-35 D^3+156 D^2-310 D+212\right)  r_0^{6 D}\cr
&&~-3 (D-3)^3 \left(55 D^4-482 D^3+1506 D^2-1992 D+952\right)  q^4 r_0^{2 (D+6)}\cr
&&~+3 (D-3) (D-2) D (5 D-11) (5 D^3-45 D^2+134 D-130)  q^2 r_0^{4 D+6}
\Big]\,,
\nn\\
&&q_{\beta_8}=(D-2)\Big[3 (D-4) (D-3)^2 (11 D^2-45 D+44) q^6 r_0^{18}\cr
&&~-(D-1) (5 D-11) \left(9 D^3-93 D^2+294 D-272\right)  r_0^{6 D}\cr
&&~-3 (D-3)^2 (D-2) \left(35 D^2-209 D+264\right)  q^4 r_0^{2 (D+6)}\cr
&&~+3 (D-3) (5 D-11)(7 D^3-59 D^2+152 D-120)  q^2 r_0^{4 D+6}
\Big]\,,
\nn\\
&&q_{2\gamma_1+\gamma_2}=8 (D-3)^2 (D-2)^2 (3 D-5) q^2 r_0^6 \big[2 (5 D-11)  r_0^{4 D}\cr
&&~+(D-4) (11 D-29) q^4 r_0^{12}-\left(5 D^2-3 D-32\right)  q^2 r_0^{2 D+6}\big]\,,
\nn\\
&&q_{\gamma_3}=16 (D-3)^2 (D-2)\Big[(D-4) \left(23 D^3-158 D^2+357 D-262\right) q^6 r_0^{18}\cr
&&~-(D-2) (D-1) (5 D-11)  r_0^{6 D}-\left(20 D^4-155 D^3+402 D^2-385 D+94\right)\cr
&&~ \times q^4 r_0^{2 (D+6)}+(D-2) (5 D-11) (D^2+D-10)  q^2 r_0^{4 D+6}
\Big]\,,
\nn\\
&&q_{\gamma_4}=16 (D-3)^2 (D-2)\Big[(D-4) \left(23 D^3-160 D^2+364 D-269\right) q^6 r_0^{18}\cr
&&~-(D-2)^2 (5 D-11)  r_0^{6 D}-\left(20 D^4-160 D^3+437 D^2-455 D+128\right)  q^4 r_0^{2 (D+6)}\cr
&&~+(5 D-11) \left(D^3-2 D^2-6 D+10\right)  q^2 r_0^{4 D+6}
\Big]\,,
\nn\\
&&q_{\alpha_3^2}=
(D-4) (D-3)^2 (735 D^8-14716 D^7+128658 D^6-641968 D^5+2000407 D^4\cr
&&~-3986580 D^3+4961176 D^2-3523424 D+1092608) q^8 r_0^{24}\cr
&&~+2 (D-3) (585 D^{10}-13437 D^9+135844 D^8-795210 D^7+2985585 D^6\cr
&&~-7535065 D^5+13066882 D^4-15699840 D^3+12985984 D^2-6998528 D\cr
&&~+1917696)  q^4 r_0^{4 (D+3)}-4 (D-3)^3 (315 D^8-5544 D^7+40354 D^6-153962 D^5\cr
&&~+311835 D^4-249666 D^3-193012 D^2+518832 D-284416)  q^6 r_0^{2 (D+9)}\cr
&&~-4 (D-3) (D-2) (3 D-7) (5 D-11) (9 D^7-138 D^6+830 D^5-2412 D^4\cr
&&~+3213 D^3-830 D^2-2032 D+1440)  q^2 r_0^{6 D+6}+(D-4) (D-2)^2\cr
&&~ \times(D-1) (3 D-7)^2 (5 D-11) (3 D^4-32 D^3+121 D^2-180 D+112)  r_0^{8 D}
\,,
\nn\\
&&q_{\alpha_3\alpha_4}=4 (D-3) (D-2)\Big[(D-4) (D-3) (525 D^7-9944 D^6+79474 D^5-348200 D^4\cr
&&~+904589 D^3-1394776 D^2+1182404 D-425144) q^8 r_0^{24}\cr
&&~+(D-4) (D-2) (D-1) (3 D-7)^2 (5 D-11) \left(3 D^2-15 D+16\right)  r_0^{8 D}\cr
&&~-(D-3) (765 D^8-14967 D^7+122222 D^6-536178 D^5+1340289 D^4\cr
&&~-1808843 D^3+934148 D^2+427988 D-510736)  q^6 r_0^{2 D+18}+(495 D^9-10224 D^8\cr
&&~+89435 D^7-427848 D^6+1194013 D^5-1859308 D^4+1153285 D^3+804836 D^2\cr
&&~-1673708 D+740928)  q^4 r_0^{4 D+12}-(3 D-7) (5 D-11) (9 D^7-126 D^6+670 D^5\cr
&&~-1620 D^4+1505 D^3+250 D^2-704 D-208)  q^2 r_0^{6 D+6}
\Big]\,,
\nn\\
&&q_{\alpha_4^2}=4 (D-3)^2 (D-2)^2\Big[(D-4) (375 D^6-7399 D^5+55738 D^4-211970 D^3+435991 D^2\cr
&&~-463767 D+200248) q^8 r_0^{24}-2 (D-1)^2 (3 D-7)^2 (5 D-11)  r_0^{8 D}-2 (225 D^7-4797 D^6\cr
&&~+39659 D^5-168283 D^4+397535 D^3-518443 D^2+336341 D-77629)  q^6 r_0^{2 D+18}\cr
&&~+(D-3) (3 D-5) \left(45 D^5-519 D^4+405 D^3+8345 D^2-25670 D+21618\right)  q^4 r_0^{4 D+12}\cr
&&~-2 (3 D-7) (5 D-11) \left(6 D^4-107 D^3+495 D^2-853 D+491\right)  q^2 r_0^{6 D+6}
\Big]\,,
\nn\\
&&q_{\alpha _3(2 \alpha _5+\alpha _6) }=
2 (D-3) (-15 D^6+215 D^5-997 D^4+935 D^3+5656 D^2-16714 D\cr
&&~ +13416) q^4 r_0^{2 (D+6)}+(15 D^7-248 D^6+1586 D^5-5000 D^4+8303 D^3-8336 D^2\cr
&&~ +8288 D-6528) q^2 r_0^{4 D+6} +(D-4) (D-3) (35 D^5-561 D^4+3429 D^3-10147 D^2\cr
&&~ +14684 D-8368) q^6 r_0^{18}-2 (D-5) (D-4) (D-2) (D-1) (3 D-7) (5 D-11) r_0^{6 D}\,,
\nn\\
&&q_{\alpha _4(2 \alpha _5+\alpha _6) }=
-(15 D^5-302 D^4+1590 D^3-2644 D^2-589 D+3466) q^4 r_0^{2 (D+6)}\cr
&&~+\left(15 D^4-635 D^3+3967 D^2-8825 D+6630\right) q^2 r_0^{4 D+6}\cr
&&~+(D-4) (25 D^4-449 D^3+2421 D^2-5243 D+4014) q^6 r_0^{18}\cr
&&~+4 (D-1) (3 D-7) (5 D-11) r_0^{6 D},\cr
&&q_{(2 \alpha _5+\alpha _6)^2 }=
12 (D-3) \left(D^2+10 D-27\right) q^2 r_0^{2 D+6}-96 (D-2)^2 r_0^{4 D}\cr
&&~+(D-4) (5 D-13) \left(D^2-16 D+31\right) q^4 r_0^{12}\ .
\eea


\begin{thebibliography}{99}

\bibitem{Kats:2006xp}
Y.~Kats, L.~Motl and M.~Padi,
``Higher-order corrections to mass-charge relation of extremal black holes,''
JHEP \textbf{12} (2007), 068, arXiv:hep-th/0606100 [hep-th].

\bibitem{Cheung:2018cwt}
C.~Cheung, J.~Liu and G.N.~Remmen,
``Proof of the weak gravity conjecture from black hole entropy,''
JHEP \textbf{10}, 004 (2018),
arXiv:1801.08546 [hep-th]


\bibitem{Cano:2019ore}
P.A.~Cano and A.~Ruip\'erez,
``Leading higher-derivative corrections to Kerr geometry,''
JHEP \textbf{05}, 189 (2019)
[erratum: JHEP \textbf{03}, 187 (2020)],
arXiv:1901.01315 [gr-qc].

\bibitem{Cano:2019oma}
P.A.~Cano, T.~Ort\'\i{}n and P.F.~Ramirez,
``On the extremality bound of stringy black holes,''
JHEP \textbf{02} (2020), 175, arXiv:1909.08530 [hep-th].

\bibitem{Cremonini:2019wdk}
S.~Cremonini, C.R.T.~Jones, J.T.~Liu and B.~McPeak,
``Higher-derivative corrections to entropy and the weak gravity conjecture in anti-de Sitter space,'' JHEP \textbf{09} (2020), 003, arXiv:1912.11161 [hep-th].

\bibitem{Ma:2020xwi}
L.~Ma, Y.Z.~Li and H.~L\"u,
``$D = 5$ rotating black holes in Einstein-Gauss-Bonnet gravity: mass and angular momentum in extremality,'' JHEP \textbf{01} (2021), 201, arXiv:2009.00015 [hep-th].

\bibitem{Agurto-Sepulveda:2022vvf}
F.~Agurto-Sep\'ulveda, M.~Chernicoff, G.~Giribet, J.~Oliva and M.~Oyarzo,
``Slowly rotating $\alpha'$-corrected black holes in four and higher dimensions,''
Phys. Rev. D \textbf{107}, no.8, 084014 (2023), arXiv:2207.13214 [hep-th].

\bibitem{Ma:2021opb}
L.~Ma, Y.~Pang and H.~L\"u,
``\ensuremath{\alpha}'-corrections to near extremal dyonic strings and weak gravity conjecture,''
JHEP \textbf{01}, 157 (2022)
%doi:10.1007/JHEP01(2022)157
[arXiv:2110.03129 [hep-th]].
%3 citations counted in INSPIRE as of 17 Apr 2023

\bibitem{Ma:2022nwq}
L.~Ma, Y.~Pang and H.~L\"u,
``Improved Wald formalism and first law of dyonic black strings with mixed Chern-Simons terms,''
JHEP \textbf{10} (2022), 142, arXiv:2202.08290 [hep-th].




\bibitem{Henriquez-Baez:2022bfi}
C.~Henr\'\i{}quez-B\'aez, J.~Oliva, M.~Oyarzo and M.I.Y.~Reyes,
``R2 corrections to the black string instability and the boosted black string,''
Phys. Rev. D \textbf{107} (2023) no.4, 044021,
arXiv:2212.07296 [hep-th].

\bibitem{Noumi:2022ybv}
T.~Noumi and H.~Satake,
``Higher derivative corrections to black brane thermodynamics and the weak gravity conjecture,''
JHEP \textbf{12} (2022), 130,
arXiv:2210.02894 [hep-th].

%\cite{Reall:2019sah}
\bibitem{Reall:2019sah}
H.S.~Reall and J.E.~Santos,
``Higher derivative corrections to Kerr black hole thermodynamics,''
JHEP \textbf{04}, 021 (2019), arXiv:1901.11535 [hep-th].

\bibitem{Cheung:2019cwi}
C.~Cheung, J.~Liu and G.N.~Remmen,
``Entropy bounds on effective field theory from rotating dyonic hlack holes,''
Phys. Rev. D \textbf{100} (2019) no.4, 046003,
arXiv:1903.09156 [hep-th].


\bibitem{Loges:2019jzs}
G.J.~Loges, T.~Noumi and G.~Shiu,
``Thermodynamics of 4D dilatonic black holes and the Weak Gravity conjecture,''
Phys. Rev. D \textbf{102} (2020) no.4, 046010
arXiv:1909.01352 [hep-th].

\bibitem{Ma:2022gtm}
L.~Ma, Y.~Pang and H.~L\"u,
``Negative corrections to black hole entropy from string theory,'' arXiv:2212.03262 [hep-th].


\bibitem{Horowitz:2023xyl}
G.~T.~Horowitz, M.~Kolanowski, G.~N.~Remmen and J.~E.~Santos,
``Extremal Kerr black holes as amplifiers of new physics,''
arXiv:2303.07358 [hep-th].






\bibitem{Jacobson:1993vj}
T.~Jacobson, G.~Kang and R.C.~Myers,
``On black hole entropy,''
Phys. Rev. D \textbf{49}, 6587-6598 (1994),
arXiv:gr-qc/9312023 [gr-qc].

\bibitem{York}
J.W.~York, Jr.,
``Role of conformal three geometry in the dynamics of gravitation,''
 Phys. Rev. Lett. 28 (1972), 1082-1085.


\bibitem{Gibbons:1976ue}
G.W.~Gibbons and S.~W.~Hawking,
``Action integrals and partition functions in quantum gravity,''
Phys. Rev. D \textbf{15}, 2752-2756 (1977).


\bibitem{Alberte:2020bdz}
L.~Alberte, C.~de Rham, S.~Jaitly and A.~J.~Tolley,
``QED positivity bounds,''
Phys. Rev. D \textbf{103}, no.12, 125020 (2021), arXiv:2012.05798 [hep-th].

\bibitem{Aalsma:2022knj}
L.~Aalsma and G.~Shiu,
``From rotating to charged black holes and back again,''
JHEP \textbf{11}, 161 (2022)
arXiv:2205.06273 [hep-th].

%\cite{Melo:2020amq}
\bibitem{Melo:2020amq}
J.F.~Melo and J.~E.~Santos,
`Stringy corrections to the entropy of electrically charged supersymmetric black holes with $\mathrm{AdS}_5\times S^5$ asymptotics,''
Phys. Rev. D \textbf{103}, no.6, 066008 (2021), arXiv:2007.06582 [hep-th].


%\cite{Bobev:2022bjm}
\bibitem{Bobev:2022bjm}
N.~Bobev, V.~Dimitrov, V.~Reys and A.~Vekemans,
%``Higher derivative corrections and AdS5 black holes,''
Phys. Rev. D \textbf{106}, no.12, L121903 (2022)
doi:10.1103/PhysRevD.106.L121903, arXiv:2207.10671 [hep-th].


%\cite{Cassani:2022lrk}
\bibitem{Cassani:2022lrk}
D.~Cassani, A.~Ruip\'erez and E.~Turetta,
``Corrections to AdS$_{5}$ black hole thermodynamics from higher-derivative supergravity,''
JHEP \textbf{11}, 059 (2022), arXiv:2208.01007 [hep-th].


\bibitem{Cano:2023dyg}
P.A.~Cano and M.~David, ``The extremal Kerr entropy in higher-derivative gravities,''
[arXiv:2303.13286 [hep-th]].














\end{thebibliography}
\end{document}